\documentclass[twocolumn,aps,pra,amsmath,amssymb,superscriptaddress,longbibliography]{revtex4-1}
\usepackage{psfrag}
\usepackage{graphicx}
\usepackage{verbatim}
\usepackage{dcolumn}
\usepackage{bm}
\usepackage{color}    
\usepackage{natbib}
\usepackage[english]{babel}
\usepackage{multirow}
\usepackage{mathtools}

\usepackage{latexsym}

\usepackage{ulem,color}

\usepackage{xspace}

\newcommand{\ket}[1]{\bigl| #1 \bigr\rangle}

\newcommand{\jy}[1]{\textcolor{black}{#1}}

\newcommand{\SA}[1]{\textcolor{black}{#1}}

\newcommand{\abs}[1]{\left\lvert #1 \right\rvert}
\DeclareMathOperator*{\erf}{erf}

\begin{document}

\title{Maximization of Extractable Randomness in a Quantum Random-Number Generator}
\author{J. Y. \surname{Haw}}
\email{jing.yan@anu.edu.au}
\affiliation{Centre for Quantum Computation and Communication Technology, Department of Quantum Science, The Australian National University, Canberra, ACT 0200, Australia}
\author{S. M. Assad}
\affiliation{Centre for Quantum Computation and Communication Technology, Department of Quantum Science, The Australian National University, Canberra, ACT 0200, Australia}
\author{A. M. \surname{Lance}}
\affiliation{QuintessenceLabs, Unit 1 Lower Ground, 15 Denison Street, Deakin, ACT, 2600 Australia}
\author{N. H. Y. \surname{Ng}}
\affiliation{Centre for Quantum Technologies, National University of Singapore, 3 Science Drive 2, 117543 Singapore}
\author{V. \surname{Sharma}}
\affiliation{QuintessenceLabs, Unit 1 Lower Ground, 15 Denison Street, Deakin, ACT, 2600 Australia}
\author{P. K. \surname{Lam}}
\affiliation{Centre for Quantum Computation and Communication Technology, Department of Quantum Science, The Australian National University, Canberra, ACT 0200, Australia}
\author{T. \surname{Symul}}
\affiliation{Centre for Quantum Computation and Communication Technology, Department of Quantum Science, The Australian National University, Canberra, ACT 0200, Australia}
\begin{abstract}
The generation of random numbers via quantum processes is an efficient and reliable method to obtain true indeterministic random numbers that are of vital importance to cryptographic communication and large-scale computer modeling. However, in realistic scenarios, the raw output of a quantum random-number generator is inevitably tainted by classical technical noise. The integrity of the device can be compromised if this noise is tampered with, or even controlled by some malicious party. To safeguard against this, we propose and experimentally demonstrate an approach that produces side-information independent randomness that is quantified by min-entropy conditioned on this classical noise. We present a method for maximizing the conditional min-entropy of the number sequence generated from a given quantum-to-classical-noise ratio. The detected photocurrent in our experiment is shown to have a real-time random-number generation rate of 14 (Mbit/s)/MHz. The spectral response of the detection system shows the potential to deliver more than 70 Gbit/s of random numbers in our experimental setup.
\end{abstract}
\date{\today}
\maketitle
\section{Introduction}
\label{sec:intro}
Randomness is a vital resource in many information and communications technology applications, such as computer simulations, statistics, gaming, and cryptography. For applications that are not concerned with the security and uniqueness of randomness, a sequence with uniformly distributed numbers mostly suffices. Such sequences can be  generated using a pseudorandom-number generator (PRNG) that works via certain deterministic algorithm. Although PRNGs can offer highly unbiased random numbers, they cannot be used for applications that require information security for two reasons: First, PRNG-generated sequences are unpredictable only under limitations of computational power, since PRNGs are inherently based on deterministic algorithms. Second, the random seeds, which are required to define the initial state of a PRNG, limit the amount of entropy in the random-number sequences they generate. This compromises the security of an encryption protocol.

For cryptographic applications \cite{Stipþeviu2011}, a random sequence
is required to be truly unpredictable and to have
maximum entropy. To achieve this, intensive efforts have been
devoted to developing high-speed hardware RNGs that generate
randomness via physical noise
\cite{xu2006compact,Sunar2007,Uchida2008,
  Kanter2009,marangon2014random}. Hardware RNGs are attractive
alternatives because they provide fresh randomness based on
physical processes that are apparently unpredictable
(i.e.\ uncorrelated with any existing information either with past
settings or side information). 
Moreover, they also provide a solution to the problem of having
insufficient entropy. Because of the deterministic nature of classical
physics, however, some of these hardware generators may be only truly
random under practical assumptions that cannot be validated. RNGs that
rely on quantum processes (QRNGs), on the other hand, can have guaranteed
indeterminism and entropy, since quantum processes are inherently
unpredictable \cite{Calude2010,Svozil2009}. Examples of such processes
include quantum phase fluctuations
\cite{Qi2010,Guo2010,Jofre2011,Xu2012a,Abellan2014}, spontaneous
emission noise \cite{Stipcevic2007,Williams2010,Liu2013}, photon
arrival times \cite{Wahl2011,Wayne2010,Ma2005}, stimulated Raman scattering
\cite{Bustard2013}, photon polarization state
\cite{Munro2007,Vallone}, vacuum fluctuations
\cite{Symul2011,Gabriel2010}, and even mobile phone cameras
\cite{Sanguinetti}. These QRNGs resolve both shortcomings of the
PRNGs. However, despite their reliance on entropy which is ultimately
guaranteed by the laws of quantum physics, measurements on quantum
systems are often tainted by classical noise. We quantify the amount
of quantum randomness to the amount of classical noise using a
quantum-to-classical-noise ratio (QCNR).  When QCNR is low, both the
quality and the security of the random sequence generated may be
compromised \cite{Gabriel2010, Ma2013, frauchiger2013true}.

To address this issue, Gabriel \textit{et al.}~\cite{Gabriel2010} took
into account potential eavesdropping on the classical noise by
considering the channel capacity of their QRNG. Their setup exhibited
a good QCNR clearance and was able to extract approximately 3 bits per
sample of guaranteed randomness out of 5 bits of digitization
(approximately 60\%). More recently, Ma \textit{et al.}~\cite{Ma2013} proposed a framework for QRNG entropy evaluation . By using \textit{min-entropy} as the quantifier for randomness, they extracted a higher rate of random bits of 6.7 bits per sample from 8 bits (approximately 84\%), where
the quantum contribution of the randomness was obtained by inferring the QCNR.

In the process of generating random bits via measuring
continuous-variable systems, an analog-to-digital converter (ADC) is
commonly used to discretize the measurement outcomes.  It has been
speculated \cite{Xu2012a} that the freedom of choosing the ADC range
could be exploited to optimize extractable randomness.  Meanwhile, in Refs.~\cite{Oliver2013,Yamazaki2013}, the choice of dynamical ADC range was
justified by experimental observations. Thus, a systematic approach to
determine the dynamical ADC range in extracting a maximum
amount of secure randomness is very much required.

In this work, we propose a new generic framework where the dynamical ADC range can be appropriately chosen to deliver maximum randomness. By quantifying randomness via min-entropy conditioned on classical noise, we show that QCNR is not the sole limiting factor in generating secure random bits. In fact, by carefully optimizing the dynamical ADC range, one can extract a nonzero amount of secure randomness even when the classical noise is larger than the quantum noise. By applying this method to our continuous-variable (CV) QRNG based on a homodyne measurement of vacuum state \cite{Symul2011}, we demonstrate that the setup is capable of delivering more than 70~Gbps of secure random bits. 

\SA{ In most QRNGs, including this work, the measurement device has to
  be calibrated and trusted.} Recently, certifiable randomness based on a
  violation of fundamental inequalities has been proposed and
  demonstrated
  \cite{Massar2010,Pironio2013,christensen2013detection}. These
  devices do not rely on the assumption of a trusted device. The
  generated bits can be certified as random based on the measured
  correlations alone and independent from the internal structure of
  the generator. However, achieving a high generation rate with such
  devices is experimentally challenging.

This paper is structured as follows. In Sec~\ref{sec:entquan}, we present the modelling of our CV QRNG and the quantification of entropy via (conditional) min-entropy. We outline the procedure of optimizing the dynamical ADC range under different operating conditions and experimental parameters. In Sec.~\ref{sec:expt}, we analyze and characterize our CV QRNG based on our framework. We then review and discuss the randomness extraction in our CV QRNG, ending with  brief concluding remarks in Sec.~\ref{sec:conc}. 

\section{Entropy Quantification}
\label{sec:entquan}
The main goal of entropy evaluation of a secure QRNG is to quantify
the amount of randomness available in the measurement outcome $M$,
conditioned upon side-information $E$, which might be accessible,
controllable, or correlated with an adversary. \SA{The concept
  of side-information-independent randomness, which includes privacy
  amplification and randomness extraction, is well established in both classical and quantum-information theory
  \cite{Shaltiel,De2012,Konig2009,Mauerer}.}
This security aspect of randomness generation started to get
considerable attention recently in the framework of QRNG
\cite{Symul2011,
  Munro2007,Ma2013,Abellan2014,Bustard2013,frauchiger2013true,Vallone,law2014quantum}. In
particular, Ref.~\cite{law2014quantum} examines the amount of randomness
extractable under various levels of characterization of the device and
power given to the adversary.

\jy{In this paper, we consider randomness as independent of classical side information, which arises from various sources of classical origin, such as technical electronic noise and thermal noise. We look at the worst-case scenario, namely that these parameters, in principle can be known by the adversary either due to monitoring or controlling of the classical noise, and, hence are untrusted. In order to guarantee the security of the random bits generated, we resort to the notion of min-entropy. This quantity is directly related to the maximum probability of observing any particular measurement outcome, hence bounding the amount of knowledge of an adversary. More specifically, the conditional min-entropy tells us the amount of (almost) uniform and independent random bits that one can extract from a biased random source with respect to untrusted parameters.} Our goal here is to achieve the maximum amount of conditional min-entropy by optimizing the measurement settings. 

\subsection{Characterization of noise and measurement}
\label{sec:charac}
\begin{figure}[t]
\includegraphics[width=0.9\columnwidth]{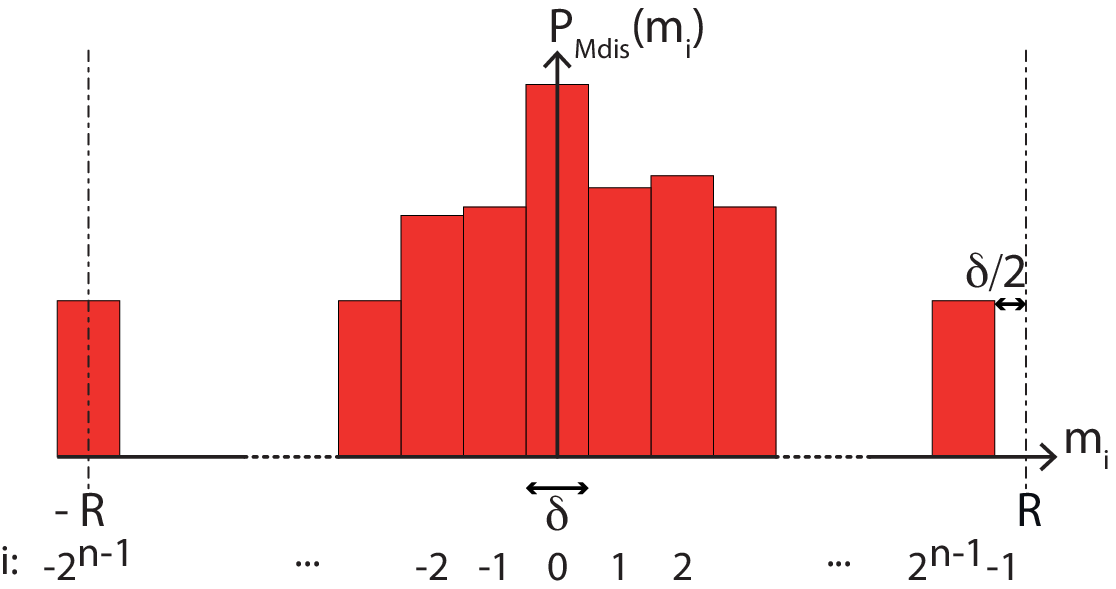}
\caption{Model of the $n$-bit ADC, with analog input in the ADC dynamical range  $[-R+\delta/2, R-3 \delta/2]$ and bin width $\delta=R/2^{n-1}$. We choose the central bin centred around $0$, the lowest bin $i_{\rm min}=-2^{n-1}$ centered around $-R$, and the highest bin $i_{\rm max}=2^{n-1}-1$ centered around $R-\delta$.}
\label{fig_ADCmodel}
\end{figure}
We first discuss the model for our CV QRNG. Following our previous
work \cite{Symul2011}, a homodyne measurement of the vacuum state is
performed. \SA{This measures $Q$, the quadrature values of the vacuum
state. The theory of quantum mechanics states that these values are
random and have a probability density function (PDF) $p_Q$ which is
Gaussian and centred at zero with variance $\sigma_M^2$. In
practice, these quadrature values cannot be measured in complete
isolation from sources of classical noise $E$. The measured signal $M$
is then $M=Q+E$. Denoting the PDF of the classical noise as $p_E$, the
resulting measurement PDF, $p_M$ is then a convolution of $p_Q$ and
$p_E$. Assuming that the classical noise follows a Gaussian
distribution centred at zero and with variance $\sigma_E^2$, the
measurement probability distribution is}
\begin{figure*}[t]
\includegraphics[width=17cm]{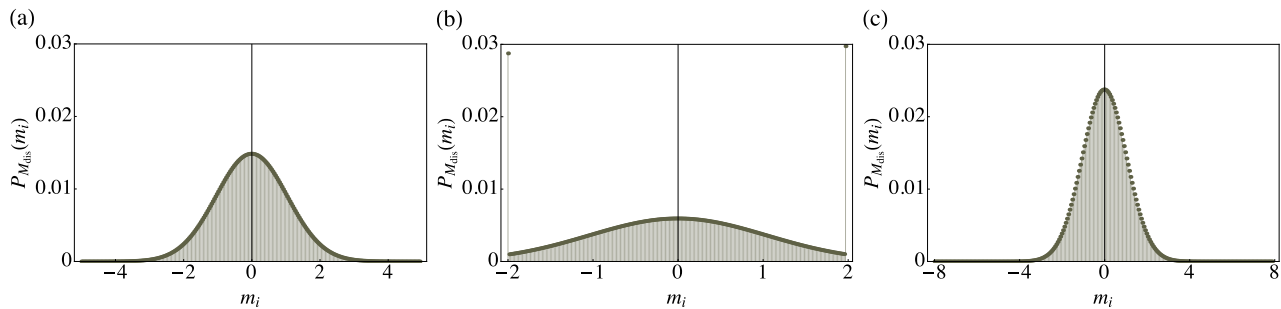}
\caption{\SA{Numerical simulations} for the measured distribution
  probabilities $P_{M_{\rm
      dis}}(m_i)$ versus quadrature values, with different dynamical ADC range parameters $R=$ (a) 5, (b) 2 and (c) 8. Without optimization, one will
  have either an oversaturated or unoccupied ADC bins, which will
  compromise both the rate and the security of the random-number
  generation. The parameters used are $n=8$ and QCNR$=10$
  $\textrm{dB}$. \SA{The quadrature values are normalized to quantum noise.}}
\label{fig:PMUop}
\end{figure*}
\begin{eqnarray}
p_M(m)=\frac{1}{\sqrt{2 \pi} \sigma_{M}} \exp \left(-\frac{m^2}{2\sigma^2_{M}}\right),
\end{eqnarray}
\SA{for $m \in M$ where the measurement variance
  $\sigma^2_M=\sigma^2_Q+\sigma^2_E$.} The ratio between the variances
of the quantum noise and the classical noise defines the QCNR, i.e.\
QCNR$=10\log_{10}(\sigma^2_Q/\sigma^2_E)$. The sampling is performed
over an $n$-bit ADC with dynamical ADC
range $[-R+\delta/2, R-3 \delta/2]$. Upon measurement, the sampled
signal is discretized over $2^n$ bins with bin width
$\delta=R/2^{n-1}$. \SA{The range is chosen so that the central bin is
centered at zero.} The resulting probability distribution of
discretized signal $M_{\rm dis}$ reads
\begin{align}
&P_{M_{\rm dis}}(m_i)\nonumber\\
&= \begin{cases}  \int_{-\infty}^{-R+\delta/2} p_{M}(m) {\rm d}m, &i=i_{\rm min}, \\
\int_{m_i-\delta/2}^{m_i+\delta/2}p_{M}(m) {\rm d}m ,&  i_{\rm min}< i< i_{\rm max}, \\
\int_{R-3 \delta/2}^{\infty} p_{M}(m) {\rm d}m, &i=i_{\rm max},
\label{eq:PHybrid}
\end{cases}
\end{align}
as shown in Fig.~\ref{fig_ADCmodel} and $m_i=\delta \times i$, where the $i$ are integers $\in \{-2^{n-1},...,2^{n-1}-1\}$. The two extreme cases $i=i_{\rm min}$ and $i=i_{\rm max}$ are introduced to model the saturation on the first and last bins of an ADC with finite input range, i.e.\ all the input signals outside $[-R+\delta/2, R-3 \delta/2]$ will be accumulated in the first and last bins. Figure \ref{fig:PMUop} shows the discretized distribution $P_{M_{\rm dis}}(m_i)$ with different $R$. We see that an appropriate choice of dynamical ADC range for a given QCNR and digitization resolution $n$ is crucial, since overestimating or underestimating the range will either lead to excessive unused bins or unnecessary saturation at the edges of the bins \cite{Oliver2013}, causing the measurement outcome to be more predictable.

However, in designing a secure CV QRNG, $R$ should not be naively
optimized over the measured distribution $P_{M_\text{dis}}(m_i)$ but over the
distribution conditioned on the classical noise. The conditional PDF
between the measured signal $M$ and the classical noise $E$, $p_{M|E}(m|e)$ is given by
\begin{eqnarray}
p_{M|E}(m|e)&=&\frac{1}{\sqrt{2 \pi (\sigma^2_M-\sigma^2_E)}} \exp
\left[-\frac{(m-e)^2}{2 (\sigma^2_M-\sigma^2_E)} \right]\nonumber\\
&=&\frac{1}{\sqrt{2 \pi} \sigma_Q} \exp
\left[-\frac{(m-e)^2}{2\sigma^2_Q} \right].
\label{eq:condpdf}
\end{eqnarray}
This is the PDF of the quantum signal shifted by the classical noise
outcome $e$. By setting $\sigma^2_Q=1$, we normalize all the relevant quantities by the quantum noise. From Eq.~\eqref{eq:PHybrid}, the discretized conditional probability distribution is, thus,
\begin{align}
&P_{M_{\rm dis}|E}(m_i|e)\nonumber\\
&= \begin{cases}  \int_{-\infty}^{-R+\delta/2}p_{M|E}(m|e) {\rm d}m , &i=i_{\rm min}, \\
\int_{m_i-\delta/2}^{m_i+\delta/2}p_{M|E}(m|e) {\rm d}m ,&  i_{\rm min}< i< i_{\rm max},\\
\int_{R-3 \delta/2}^{\infty} p_{M|E}(m|e) {\rm d}m , &i=i_{\rm max}.  
\end{cases}
\label{eq:PcondHybrid}
\end{align}
With these, we are now ready to discuss how $R$ should be chosen under two different definitions of min-entropy, namely worst-case min-entropy and average min-entropy. 
\subsection{Worst-case conditional min-entropy}
\label{sec:Cond_min}
\begin{figure*}[t]
\includegraphics[width=15cm]{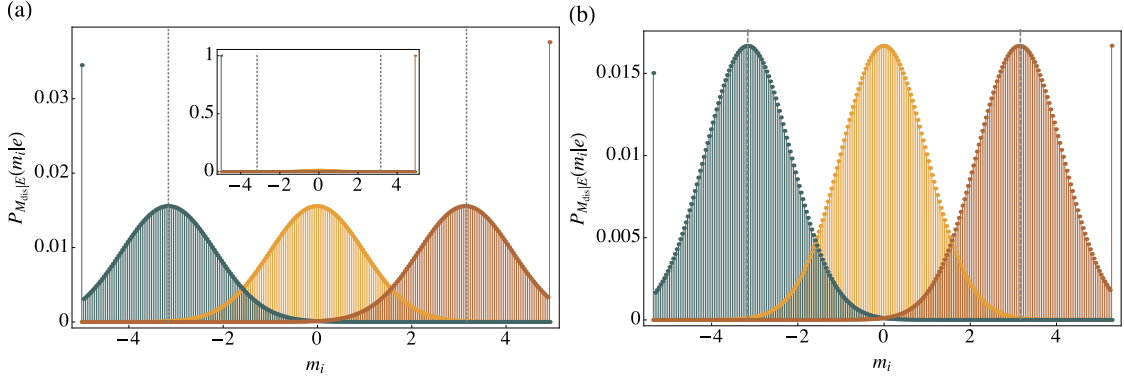}
\caption{\SA{Numerical simulations of:} (a) conditional probability
  distributions $P_{M_{\rm dis}|E}(m_i|e)$, with
  $e=\{-10\sigma_E,0,10\sigma_E\}$ (from left to right) and
  $R=5$. Without optimizing $R$, when $e=\pm10\sigma_E$, saturations in
  the first and last bins affect the maximum of the conditional
  probability distribution. Inset: $P_{M_{\rm dis}|E}(m_i|e)$, with
  $e=\{-100\sigma_E,0,100\sigma_E\}$ (from left to right). Unbounded
  classical noise will lead to zero randomness due to the
  oversaturation of dynamical ADC. (b) Optimized $P_{M_{\rm
      dis}|E}(m_i|e)$, with $e=\{-10\sigma_E,0,10\sigma_E\}$ (from
  left to right). From Eq.~\eqref{eq:wcHminOpE}, the optimal $R$ is
  chosen to be $5.35$. The saturations do not exceed the maximum of the
  conditional probability distribution whenever $-10\sigma_E\leq e\leq
  10\sigma_E$. The parameters are $n=8$, QCNR$=10$ $\textrm{dB}$. Dashed lines indicate $m_i=\pm10\sigma_E$.  \SA{The quadrature values are normalized to vacuum noise.}}
\label{PMECB}
\end{figure*}
The min-entropy for variable $X$ with distribution $P_X(x_i)$, in unit of bits, is defined as \cite{Konig2009,dodis2008fuzzy}:
\begin{eqnarray}
H_{\rm min}(X)=-\log_2\left[\max_{x_i \in X} P_X(x_i)\right].
\label{eq:Hmindef}
\end{eqnarray} 
Operationally, this corresponds to entropy associated with the maximum guessing probability for an eavesdropper about $X$. It also tells us about how much (almost) uniform randomness can be extracted out of the distribution $P_X(x_i)$. To obtain a lower bound for the randomness in our entropy source, we first look into the worst-case min-entropy conditioned on classical side information $K$, which is defined as \cite{renner2008security} 
\begin{equation}
\begin{aligned}
& H_{\rm min}(X|K)\\
&=-\log_2\left[ \max_{k_j \in {\rm supp}(P_K)} \max_{x_i \in X} P_{X|K}(x_i|k_j)\right],\\
 \label{eq:Hmincondef}
\end{aligned}
\end{equation}
where the support ${\rm supp}(f)$ is the set of values $x_i$ such that $f(x_i) > 0$. In the case of Gaussian distributions, the support of the probability distribution will be $\mathbb{R}$. Following Eq.~\eqref{eq:PcondHybrid}, upon discretization of the measured signal $M$, the worst-case min-entropy conditioned on classical noise $E$ is
\begin{equation}
H_{\rm min}(M_{\rm dis}|E)
=-\log_2\left[ \max_{e \in \mathbb{R}} \max_{m_i \in M_{\rm dis}} P_{M_{\rm dis}|E}(m_i|e)\right].
\label{eq:wcHmin}
\end{equation}
Here we assumed that from the eavesdropper's perspective, the classical noise is known fully with arbitrary precision. Performing the integration in Eq.~\eqref{eq:PcondHybrid}, the maximization over $M_{\rm dis}$ in Eq.~\eqref{eq:wcHmin} becomes
\begin{align}
 &\max_{m_i \in M_{\rm dis}} P_{M_{\rm dis}|E}(m_i|e)\nonumber\\
= &\max \begin{cases} \frac{1}{2}\left[1-\erf\left(\frac{e+R-\delta/2}{\sqrt{2}}\right)\right], \\
\erf\left(\frac{\delta}{2 \sqrt{2}}\right),\\
\frac{1}{2}\left[\erf\left(\frac{e-R+3 \delta/2}{\sqrt{2}}\right)+1\right], 
\end{cases}
\label{eq:PcondHybridProb}
\end{align}
where $\erf(x)=2/\sqrt{\pi} \int_0^x e^{-t^2} dt$ is the error function. We note that we have $ \max_{e \in \mathbb{R}} \max_{m_i \in M_{\rm dis}} P_{M_{\rm dis}|E}(m_i|e)=1$, achieved when $e\rightarrow -\infty$ or $e \rightarrow \infty$. This results in $H_{\rm min}(M_{\rm dis}|E)=0$ [see inset of Fig.~\ref{PMECB} (a)]. Indeed it is intuitive to see that in the case where the classical noise $e$ takes on an extremely large positive value, the outcome of $M_{\rm dis}$ is almost certain to be $m_{i_{\max}}$ with large probability. However, this scenario happens with a very small probability. Hence for practical purposes, one can bound the maximum excursion of $e$, for example $-5 \sigma_E \leq e \leq 5\sigma_E$, which is valid for $99.9999\%$ of the time. With this bound on the classical noise, we now have
\begin{align}
&\max_{e \in [e_{\rm min},e_{\rm max}]}  \max_{m_i \in M_{\rm dis}} P_{M_{\rm dis}|E}(m_i|e)\nonumber\\
&= \max \begin{cases} \frac{1}{2}\left[1-\erf\left(\frac{e_{\rm min}+R-\delta/2}{\sqrt{2}}\right)\right], \\
\erf\left(\frac{\delta}{2 \sqrt{2}}\right),  \\
\frac{1}{2}\left[\erf\left(\frac{e_{\rm max}-R+3 \delta/2}{\sqrt{2}}\right)+1\right],  \label{eq:PcondHybridProbMax}
\end{cases}
\end{align}
and when $e_{\rm min}=e_{\rm max}$,
\begin{widetext}
\begin{equation}
H_{\rm min}(M_{\rm dis}|E)=
-\log_2 \left[  \max \left\{
\frac{1}{2}\left[\erf\left(\frac{e_{\rm max}-R+3 \delta/2}{\sqrt{2}}\right)+1\right];\erf\left(\frac{\delta}{2 \sqrt{2}}\right) \right\}\right],
\label{eq:wcHminOp}
\end{equation}
\end{widetext}
which can be optimized by choosing $R$ such that
\begin{equation}
\frac{1}{2}\left[\erf\left(\frac{e_{\rm max}-R+3 \delta/2}{\sqrt{2}}\right)+1\right]=\erf\left(\frac{\delta}{2 \sqrt{2}}\right). 
\label{eq:wcHminOpE}
\end{equation}
This optimized worst-case min-entropy $H_{\rm min}(M_{\rm dis}|E)$ is directly related to the extractable secure bits that are independent of the classical noise. As shown in Fig.~\ref{PMECB}(a), when Eq.~\eqref{eq:wcHminOp} is not optimized with respect to $R$, the saturation in the first (last) bin for $e_{\rm min/max}=\pm10\sigma_E$ becomes the peaks of the conditional probability distribution, hence compromising the attainable min-entropy. By choosing the optimal value for $R$ via Eq.~\eqref{eq:wcHminOpE}, as depicted in Fig.~\ref{PMECB}(b), the peaks at the first and last bins will always be lower than or equal to the probability within the dynamical range. Thus, by allowing the dynamical ADC range to be chosen freely, one can obtain the lowest possible conditional probability distribution, and hence produce the highest possible amount of secure random bits per sample for a given QCNR and $n$-bit ADC. Equation \eqref{eq:wcHminOp} can be further generalized to take into account the direct current (dc) offset of the device $\Delta$, which can be due to an intrinsic offset of the electronic signal  or even a deliberate constant offset induced by the eavesdropper over the sampling period [see Appendix \ref{app:dc}]. 

\begin{figure}[t]
\includegraphics[width=1\columnwidth]{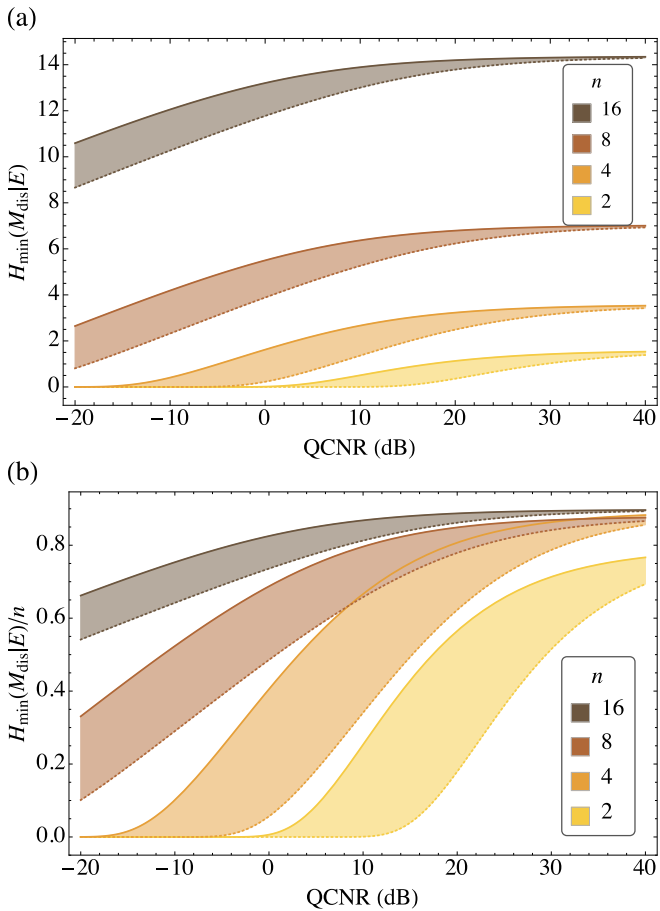}
\caption{(a) Optimized $H_{\rm min}(M_{\rm dis}|E)$ and (b) normalized $H_{\rm min}(M_{\rm dis}|E)$ as a function of QCNR for different $n$-bit ADCs. Shaded areas: $5\sigma_E\leq |e +\Delta| \leq 20\sigma_E$. The extractable bits are robust against the excursion of the classical noise, especially when the QCNR is large. A nonzero amount of secure randomness is extractable even when the classical noise is larger than the quantum noise. The extractable secure randomness per bit increases as the digitization resolution $n$ is increased.}
\label{fig:HminWC}
\end{figure}
\begin{figure}[t]
\includegraphics[width=1\columnwidth]{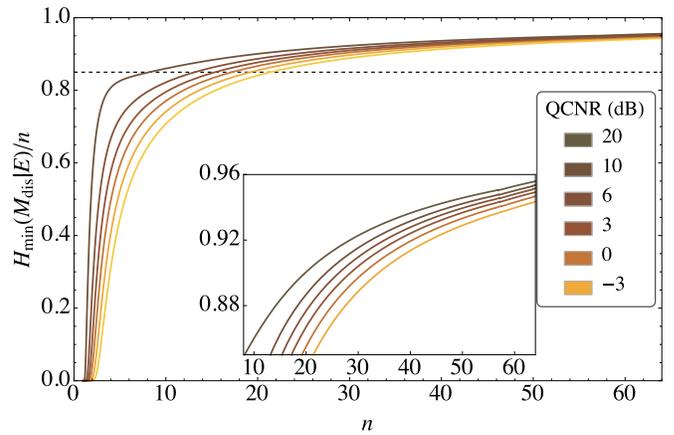}
\caption{Normalized worst-case conditional min-entropy $H_{\rm
    min}(M_{\rm dis}|E)$ as a function of $n$-bit ADC for different
  QCNR values. $|\Delta|=0$ and $|e|\leq 5\sigma_E$. The interplay
  between the QCNR and digitization resolution $n$ is shown, where one can
  improve the rate of secure randomness per bit either by improving
  the QCNR or increasing $n$. Inset: Zoom in for $H_{\rm min}(M_{\rm
    dis}|E)/n \geq 0.85$ (dashed line). Even when the classical noise
  is more dominating compared to the quantum noise (QCNR$=-3$ dB), 85 \% of the randomness per bit can be recovered by having at least approximately 22 bits of digitization.}
\label{NHminwn}
\end{figure}

In Fig.~\ref{fig:HminWC}(a), we show the extractable secure random
bits for different digitization $n$ under the confidence interval of
$5\sigma_E\leq|e+\Delta|\leq 20\sigma_E$. At the high QCNR regime, the
classical noise contribution does not compromise the extractable bits
too much. As the classical noise gets more and more comparable to the
quantum noise, although more bits have to be discarded, one can still
extract a decent amount of secure random bits. More surprisingly, even
if the QCNR goes below 0, that is, classical noise becomes larger than
quantum noise, in principle, one can still obtain a nonzero amount of
random bits that are independent of classical noise. From
Fig.~\ref{fig:HminWC}(b), we notice the extractable secure randomness per
bit increases as we increase the digitization resolution $n$. This interplay
between the digitization resolution $n$ and QCNR is further explored in
Fig.~\ref{NHminwn}, where normalized $H_{\rm min}(M_{\rm dis}|E)$ is
plotted against $n$ for several values of QCNR. We can see that for
higher ratios of quantum-to-classical-noise, a lesser amount of
digitization resolution is required to achieve a certain value of secure
randomness per bit. In other words, even if QCNR cannot be improved
further, one can achieve a higher ratio of secure randomness per bit
simply by increasing $n$.
\subsection{Average conditional min-entropy}
As described in Section \ref{sec:Cond_min}, without a bound on the range of classical noise, one cannot extract any secure randomness. However, if we assume that an adversary can only listen to, but has no control over the classical noise, 
we can estimate the average chance of successful eavesdropping with
the {\it average} guessing probability of $M_{\rm dis}$ given $E_{\rm dis}$  \cite{Konig2009,dodis2008fuzzy,Pironio2013},  
\begin{equation}
\begin{aligned}
&P_\mathrm{guess}(M_{\rm dis}|E_{\rm dis})\\
&=\left[ \sum_{e_j \in E_{\rm dis}} P_{E_{\rm dis}}(e_j)\max_{m_i \in M_{\rm dis}} P_{M_{\rm dis}|E_{\rm dis}}(m_i|e_j)\right],
\end{aligned}
\label{eq:pguess}
\end{equation}
which denotes the probability of correctly predicting the value of discretized measured signal $M_{\rm dis}$ using the optimal strategy, given access to discretized classical noise $E_{\rm dis}$. Here $P_{E_{\rm dis}}(e_j)$ is the discretized probability distribution of the classical noise. The extractable secure randomness from our device is then quantified by the average conditional min-entropy
\begin{equation}
\bar{H}_{\rm min}(M_{\rm dis}|E_{\rm dis})=-\log_2 P_\mathrm{guess}(M_{\rm dis}|E_{\rm dis}).
\label{eq:aveHmin}
\end{equation}
Here, we again assume that the eavesdropper can measure the full spectrum of the classical noise, with arbitrary precision. This gives the eavesdropper maximum power, including an infinite ADC range $R_e\rightarrow \infty$ and infinitely small binning $\delta_e\rightarrow 0$. As detailed in Appendix \ref{app:bin}, under these limits, Eq.~\eqref{eq:aveHmin} takes the form of
\begin{equation}
\begin{aligned}
&\bar{H}_{\rm min}(M_{\rm dis}|E)\\
&=\lim_{\delta_e\rightarrow 0} \bar{H}_{\rm min}(M_{\rm dis}|E_{\rm dis})\\
&=-\log_2\left[ \int^\infty_{-\infty} P_E(e)\max_{m_i \in M_{\rm dis}} P_{M_{\rm dis}|E}(m_i|e){\rm d}e\right].
\end{aligned}
\label{eq:pguessMVN}
\end{equation}
\begin{table}[t]
\centering
\caption{Optimized $\bar{H}_{\rm min}(M_\mathrm{dis}|E)$ (and $R$) for 8- and 16-bit ADCs}
\begin{tabular} {l c c}
\hline\hline
 QCNR (dB) & $n=8$  &$n=16$  \\ [0.5ex]
\hline
$\infty$ & 7.03 (2.45) & 14.36 (3.90)\\
20 &  6.93 (2.59) & 14.28 (4.09)\\
10 &  6.72 (2.93)&  14.11 (4.55) \\
0 & 6.11 (4.33)  & 13.57 (6.48)\\
-$\infty$ & 0 & 0\\
\hline
\end{tabular}
\label{tab:aveHmin}
\end{table}
The full expression of Eq.~\eqref{eq:pguessMVN} is shown in Eq.~\eqref{eq:pguessMVNX}. The optimized result for the average min-entropy $\bar{H}_{\rm min}(M_{\rm dis}|E)$ with the corresponding dynamical ADC range $R$ is depicted in Table \ref{tab:aveHmin}. Similar to worst-case min-entropy scenario in Sec.~\ref{sec:Cond_min}, one can still obtain a significant amount of random bits even if the classical noise is comparable to quantum noise. On the contrary, a conventional unoptimized QNRG requires high operating QCNR to access the high-bitrate regime. When QCNR$\rightarrow\infty$, the measured signal does not depend on the classical noise and the result coincides with that of the worst-case conditional min-entropy. In fact, the worst-case conditional min-entropy [Eq.~\eqref{eq:wcHmin}] is the lower bound for the average conditional min-entropy [Eq.~\eqref{eq:aveHmin}]. In the absence of side-information $E$, both entropies will reduce to the usual min-entropy Eq.~\eqref{eq:Hmindef} \cite{Pironio2013}. Compared to the worst-case min-entropy, the average conditional min-entropy is more robust against degradation of QCNR; hence, it allows one to extract more secure random bits for a given QCNR. This is expected, \jy{since in this case, we do not allow the eavesdropper to influence our device,} which is a valid assumption for a trusted laboratory.
\section{Experimental Implementation}
\label{sec:expt}
\subsection{Physical setup and characterization}
\label{sec:physical}
\begin{figure}[t]
\includegraphics[width=1\columnwidth]{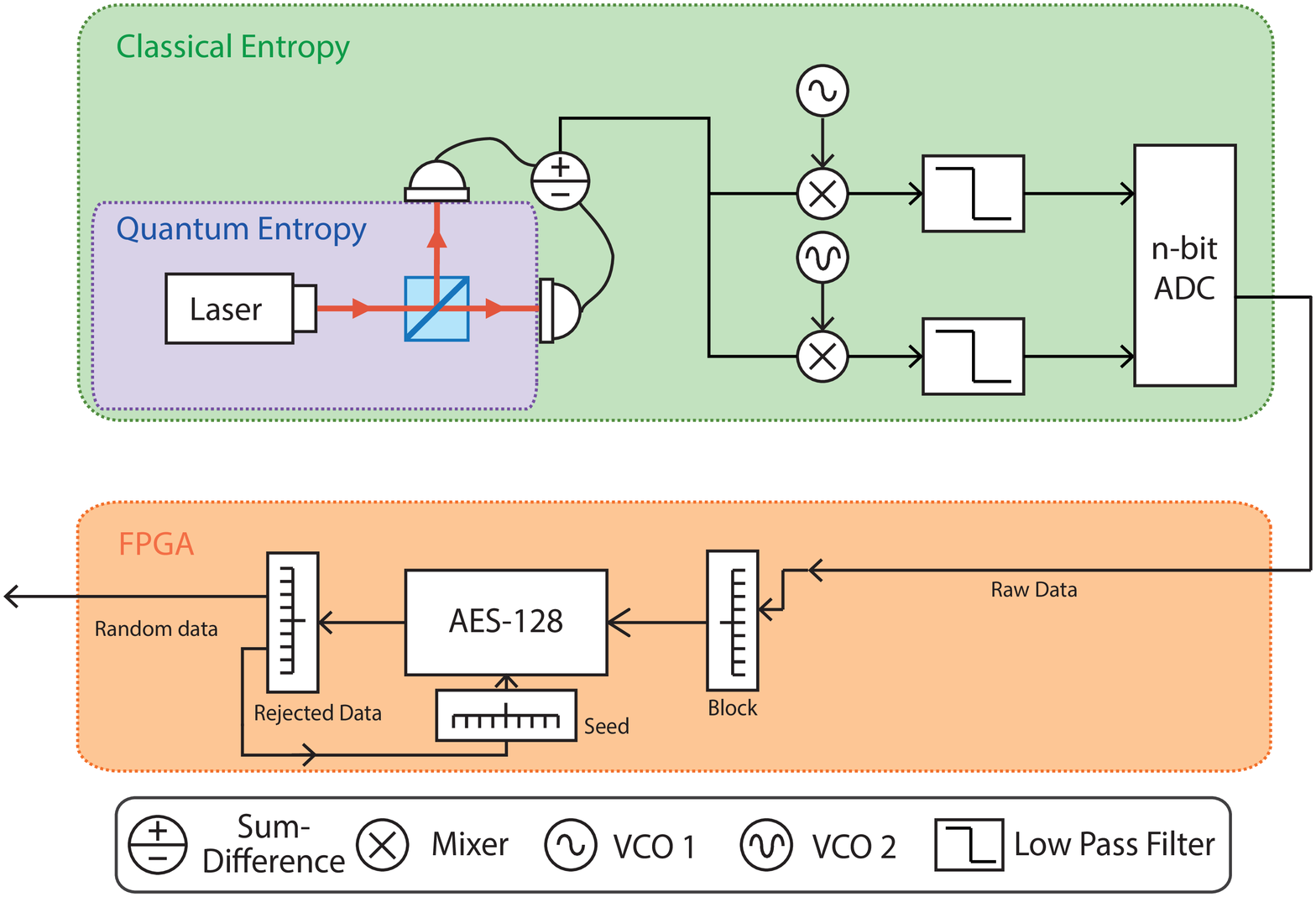}
\caption{Schematic setup of CVQRNG, where a continuous-variable
  homodyne detection is performed on the quantum vacuum state,
  followed by mixing down at 1.375 GHz and 1.625 GHz. The mixing
  signals are generated by voltage-controlled oscillators. The
  dynamical ADC range of the ADC is chosen appropriately according to
  the QCNR and ADC digitization resolution $n$ to maximize the extractable
  randomness. The raw output, which consists of both quantum and classical contributions, will be postprocessed by field-programmable gate array. A cryptographic hashing function (AES-128) is applied to extract secure randomness quantified by conditional min-entropy.}
\label{fig:schema}
\end{figure}
As depicted in Fig.~\ref{fig:schema}, our CVQRNG setup consists
  of a homodyne detection of the quantum vacuum state followed
  by post-processing. A 1550-nm fibre-coupled laser (NP Photonic Rock)
  operating at 60 mW serves as the local oscillator of the
  homodyning setup. \SA{This local oscillator is sent into one port of a
  50:50 beam splitter, while the other one is physically blocked and
  serves as the vacuum input. The outputs are then optically coupled
  to a pair of balanced photodetectors with 30 dB of common-mode
  rejection. The intensity of the output ports are recorded over a
  detection bandwidth of 3 GHz. Since the local oscillator's amplitude
  $\alpha$ is significantly larger than the quantum vacuum
  fluctuation, the difference of the photocurrents from the pair of
  detectors is proportional to $\abs{\alpha}X_v$, where $X_v$ is the
  quadrature amplitude of the vacuum state. Hence, the
  contribution of quantum noise is essentially amplified via the
  balanced homodyne detection.}

\begin{figure}[t]
\includegraphics[width=1\columnwidth]{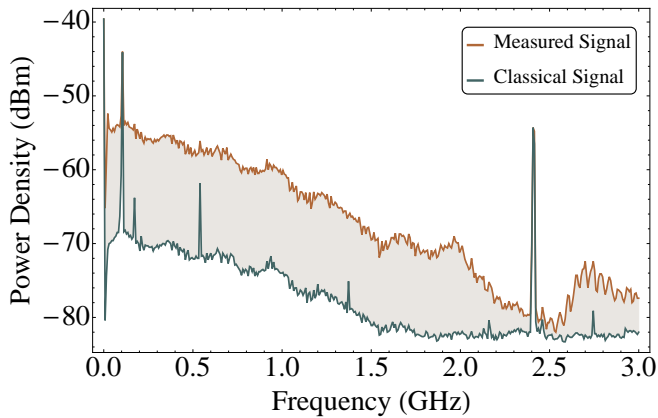}
\caption{Spectral power density from the CV-QRNG. The measured signal
  is mixed down at 1.375 GHz and 1.625 GHz (dashed lines), where the
  laser is shot-noise limited and far from low-frequency technical
  noise. The QCNR clearances are about 13 dB for both channels,
  which are sampled at 250 MSamples per second. The shaded region
  between the measured signal and classical noise indicates the
  available quantum randomness in our broadband 3-GHz photocurrent
  detectors, with an average QCNR of approximately 10 dB. \SA{The peaks in the
    classical signal are due to technical noise and pickup signals from
    radio stations. The peak at 2.4 GHz is due to the Wi-Fi transmissions.} The resolution and video bandwidth are both 1 MHz.}
\label{Sp1}
\end{figure}
\SA{In order to sample the vacuum field at the spectral range where
  technical noise is less significant and where the laser is shot-noise
  limited (see Fig.~\ref{Sp1}), the electronic output is split and
  mixed down at 1.375 GHz and 1.625 GHz. Low-pass filters with cutoff
  frequency at 125 MHz are used to minimize the correlations between the sampling points
  \cite{shen2010practical} before digitizing with an appropriately chosen dynamical
  ADC range parameter $R$.} The measured signal from two sidebands
  (channel 0, 1.25-1.50 GHz), and (channel 1, 1.50-1.75 GHz) are
  recorded using two 16-bit ADCs (National Instruments 5762) at 250
  MSamples per second. Finally, the data processing is performed
  using a National Instruments field-programmable gate array.

The average QCNR clearances for channel 0 (ch 0) and channel 1 (ch 1) are $13.52$  and $13.32$ dB, respectively. Taking into account the intrinsic dc offsets, which is $-0.02\sigma_Q$ for both channels, we quantify our conditional min-entropies using the method described in Sec. \ref{sec:entquan}. For our ADC with 16 bits of digitization, the worst-case conditional min-entropies are $13.76$ bits (ch 0) and $13.75$ bits (ch 1), while the average conditional min-entropies are $14.19$ bits for both channels. 
Here, by assuming that the eavesdropper cannot manipulate the classical noise, we evaluate our entropy with average conditional min-entropy and set $R$ as $4.32 \sigma_Q$ according to Eq.~\eqref{eq:pguessMVNXD}. 
\subsection{Upper bound of extractable min-entropy}
\label{sec:UB}
The extractable randomness of our QRNG is limited by the sampling rate and the digitization resolution, which is defined by Nyquist's theorem on maximum data rate $C$,
\begin{equation}
C=2H \log_2 V,
\label{eq:nyquist}
\end{equation}
where $H$ is the bandwidth of the spectrum and $V=2^n$ is the
quantization level for digitization resolution $n$. For our 16-bit ADC, the
shot-noise-limited and technical-noise-free bandwidth is around 2.5
GHz out of 3 GHz. With an average of 10 dB of QCNR clearance, one can
extract 14.11 bits out of 16 bits (Table
\ref{tab:aveHmin}). Putting these values into
  Eq. \eqref{eq:nyquist}, with a fast enough ADC, we can \SA{potentially}
  extract up to 70 Gbit/s random bits out of our detectors.

The maximum bitrate is ultimately upper bounded by the
  photon number within a given detection time window. In our setup, a
  1550-nm fibre-coupled laser with power of 60 mW and detection
  bandwidth of 3 GHz is used. This corresponds to a mean of $1.6 \times 10^8$
  photons per sampling. Given a perfect photon-number-resolving
  detector, the maximum min-entropy \SA{is given by $-\log_2(1/\sqrt{2\pi\times 1.6 \times 10^8}) \approx 14.9$ bits (see Appendix \ref{app:hmin_limit})}. In
  principle, one can send more power to extract more random bits, however, this bound can increase only logarithmically with laser
  intensity.
\subsection{Randomness extraction}
\label{sec:RE}
It is commonly the case that QRNGs are not ideal sources of randomness, in the sense that the distribution is often biased, while uniform randomness is required for application purposes. In our situation, the quantum vacuum state measured by our CV QRNG exhibits a Gaussian distribution. To generate ideal randomness, postprocessing of the raw outputs is necessary to produce shorter, yet almost uniformly distributed random strings. \textit{Ad hoc} algorithms such as the Von Neumann extractor, XOR corrector, and least significant bit operation are widely used \cite{Durt2013,Liu2013,Uchida2008,Yamazaki2013,Oliver2013,Williams2010}. These methods, although simple in practice, might fail to produce randomness at all if non-negligible correlations exist among the raw bits \cite{Barak}.

From an information-theoretic standpoint, universal hashing
functions are desirable candidates for randomness extraction
\cite{Shaltiel,Ma2013}. These functions act to recombine bits within a
sample according to a randomly chosen seed, and map
them to truncated, almost uniform random strings. \SA{They constitute a
strong extractor which implies that the seed can be reused
without sacrificing too much randomness.} In recent development of QRNGs
\cite{Bustard2013,Ma2013,Nie2014,Vallone,frauchiger2013true}, they have been used to
construct hashing functions such as the Toeplitz-hashing matrix. These constructions require a long (but
reusable) seed \cite{barak2011leftover}. A different implementations of an
information-theoretic randomness extractor, the Trevisan's extractor,
\cite{De2012,Mauerer,Ma2013} has also received considerable attention. This
particular construction of a strong extractor has been proven secure
against quantum side-information, and, furthermore, it requires a relatively
short seed. Despite so, the complexity of the algorithm imposes a very
stringent limit on the extraction speed (0.7 kb/s \cite{Ma2013} and
150 kb/s \cite{Mauerer}).

Another attractive alternative for secure randomness extraction
  are cryptographic hashing functions
  \cite{dodis2004randomness,Hugo2010,Yvonne2009,chevassut2005}. \SA{While
  these cryptographic hashing functions are not
  information-theoretically proven to be secure,} they are still
  suited for many cryptographic applications and settings where the
  adversary is assumed to be computationally bounded. The reason for
  utilizing them over universal hashing functions is that they can have
  high throughput due to efficient hardware
  implementation. Previously, cryptographic hashing extractors have
  been deployed in \cite{Abellan2014,Gabriel2010,Jofre2011,Wayne2010},
  with functions such as SHA-512 and Whirlpool. Most of the
  implementations keep exactly min-entropy number of bits, which might
  not be fully secure (see Appendix \ref{sec:lhl}).

Here, we demonstrate randomness extraction with the Advanced Encryption Standard (AES) \cite{AEShash} cryptographic hashing algorithm of 128 bits (see Appendix \ref{sec:ac}). Since a detailed cryptoanalysis of our framework is nontrivial and beyond the scope of this paper, we keep only half of our conditional min-entropy \cite{Barker2012a,dodis2004randomness} to obtain an almost perfectly uniform output. The final real-time guaranteed-secure random number generation rate of our CV QRNG is $3.55$  Gbps. If all the available bandwidth from our detector (approximately 2.5 GHz) can be sampled, \jy{with sufficient resources}, we can achieve up to 35 Gbit/s (cf.~Sec.~\ref{sec:UB}). This corresponds to a rate of 14 Mbps/MHz in term of bits per bandwidth. Our random numbers consistently pass the standard statistical tests (NIST \cite{rukhin2010statistical}, DieHard \cite{marsaglia1998diehard}) and the results are available on \jy{the Australian National University} Quantum Random Number Server (https://qrng.anu.edu.au). 
\section{Concluding Remarks}
\label{sec:conc}
In this work, we propose a generic framework for secure random-number generation, taking into account the existence of classical side information, which, in principle could be manipulated or predicted by an adversary. If the adversary is assumed to have access to the classical noise, for example, the detectors' noise can be originating from preestablished values, the worst-case conditional min-entropy should be used to quantify the available secure randomness. Meanwhile, if we restrict the third party to passive eavesdropping, one can use the average conditional min-entropy instead to quantify extractable randomness. By treating the dynamical ADC range as a free parameter, we show that QCNR is not the sole decisive factor in generating secure random bits. Surprisingly, one can still extract a nonzero amount of secure randomness even when the classical noise is comparable to the quantum noise. This is done simply by optimizing the dynamical ADC range via conditional min-entropies. Such an approach not only provides a rigorous justification for choosing the suitable ADC parameter, but also largely increases the range of QCNR for which true randomness can be extracted, thus relaxing the condition of high QCNR clearance in conventional CV QRNGs. We also notice that we can increase the min-entropy per bit simply by increasing the number of digitization bits. 
We apply these observations to analyze the amount of randomness produced by our CV QRNG setup. Efficient cryptographic hashing functions are then deployed to extract randomness quantified by average conditional min-entropy. 

We note several possible extensions of our work. For instance, one can apply entropy smoothing \cite{Tomamichel2011a,renner2008security} on the worst-case min-entropy to tighten the analysis. Our framework can also be generalized to encapsulate potential quantum side information by considering the analysis described in Ref.~\cite{frauchiger2013true}. A detailed cryptoanalysis of our framework can also increase the final throughput of the QRNG \cite{Yvonne2009}. Last, a hybrid of an information-theoretic provable and cryptographic randomness extractor is also an interesting avenue to be explored in the construction of a high-speed, side-information (classical and quantum) proof QRNG \cite{barak2011leftover}.

To conclude, this work allows the maximization of extractable high-quality randomness without compromising both the integrity and the speed of a QRNG. In fact, within our framework, when the QRNG is appropriately calibrated, the generated random numbers are secure even if the electronic noise is fully known. This is of practical importance, given the fact that QRNGs play a decisive role in the implementation of cryptographic protocols such as quantum key distribution. From a practical point of view, our method also relaxes the QCNR requirement on the detector, thus allowing QRNGs that are more cost effective and smaller in size. As such, we believe that our work paves the way towards a reliable, high bitrate, and environmentally-immune QRNG \cite{gisin2010quantum} for information security.

\SA{\textit{Note added}.- We note several recent papers considering the security aspects of QRNGs
\cite{Lougovski2014,stipvcevic2014post,lunghi2014self,canas2014experimental}. A related work by Mitchell \textit{et al.}~\cite{Mitchell2015} was published recently. Similar to our work, a lower bound
of the average min-entropy was established by assuming the worst-case behavior of various untrusted experimental
noise and errors. The study asserts that high-bitrate randomness generation with strong randomness
guarantee remains possible even under paranoid analysis, which supports our conclusion.}

\section*{Acknowledgement}
This research was conducted by the Australian Research Council Centre of Excellence for Quantum Computation and Communication Technology (project number CE110001027). N. N. is funded by the National Research Foundation
Competitive Research Programme “Space Based Quantum Key Distribution".
\appendix
\section{Optimized conditional min-entropy with dc offset}
\label{app:dc}
\begin{figure}[h!]
\includegraphics[width=1\columnwidth]{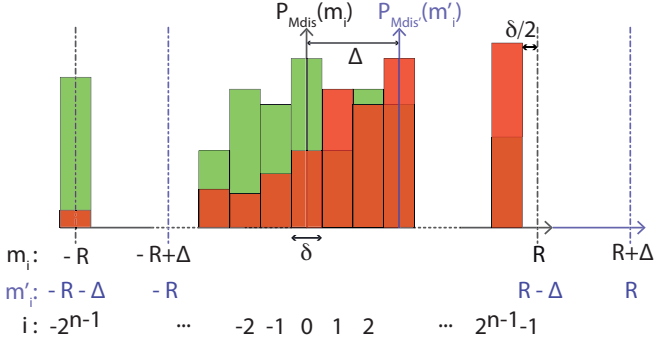}
\caption{Model of the $n$-bit ADC, with analog input in the ADC dynamical range  $[-R+\delta/2, R-3 \delta/2]$ and bin width $\delta=R/2^{n-1}$. Offset of the distribution is modeled by another reference frame $m'$ centered at offset $\Delta$. In the original frame $m$, the lowest and highest bins are now centered around $-R-\Delta$ and $R-\delta-\Delta$.}
\label{fig_ADCmodelDC}
\end{figure}
In a realistic scenario, the mean of the measured signal's probability distribution is often nonzero. It is possible that such an offset might be induced by a malicious party over the sampling period. The model is depicted in Fig.~\ref{fig_ADCmodelDC}, where the offset $\Delta$ of the distribution is captured by another reference frame $m'$ centered at $\Delta$. In this model, Eq.~\eqref{eq:PcondHybrid} can now be rewritten as
\begin{align}
&P^{(\Delta)}_{M_{\rm dis}|E}(m_i|e)= \nonumber\\
&\begin{cases}  \int_{-\infty}^{-R-\Delta+\delta/2} p_{M'|E}(m'|e) {\rm d}m', &i=i_{\rm min}, \\
\int_{m'_i-\Delta-\delta/2}^{m'_i-\Delta+\delta/2}p_{M'|E}(m'|e) {\rm d}m' ,& i_{\rm min}<i<i_{\rm max},\\
\int_{R-3 \delta/2-\Delta}^{\infty} p_{M'|E}(m'|e) {\rm d}m', &i=i_{\rm max.}\label{eq_PcondHybridx}
\end{cases}
\end{align}
\begin{table*}[t]
\caption{Optimized $H_{\rm min}(M_{\rm dis}|E)$ (and $R$) for an 8-bit ADC}
\centering
\begin{tabular} {l c c c c c}
\hline\hline
 \multirow{2}{*}{QCNR (dB)}& \multicolumn{5}{c}{$|e+\Delta|$} \\
  \cline{2-6}
  & $0$&  $5\sigma_E$ &  $10\sigma_E$ & $15\sigma_E$ &  $20\sigma_E$\\ [0.5ex]
\hline
$\infty$ & 7.03 (2.45) & 7.03 (2.45) & 7.03 (2.45) & 7.03 (2.45) & 7.03 (2.45)\\
 20  &  & 6.79 (2.90)& 6.58 (3.35)& 6.40 (3.81)& 6.23 (4.27)\\
 10  & & 6.37 (3.88)& 5.91 (5.35)& 5.55 (6.85)& 5.26 (8.36)\\
 0  & & 5.50 (7.10)& 4.75 (11.92)& 4.25 (16.82)& 3.88 (21.75)\\
-$\infty$ & & 0 &0 & 0& 0\\
\hline
\end{tabular}
\label{tab_8bits}
\end{table*}
\begin{table*}[t]
\caption{Optimized $H_{\rm min}(M_{\rm dis}|E)$ (and $R$) for a 16-bit ADC}
\centering
\begin{tabular} {l c c c c c}
\hline\hline
 \multirow{2}{*}{QCNR (dB)}& \multicolumn{5}{c}{$|e+\Delta|$} \\
  \cline{2-6}
  & $0$&  $5\sigma_E$ &  $10\sigma_E$ & $15\sigma_E$ &  $20\sigma_E$\\ [0.5ex]
\hline
$\infty$ & 14.36 (3.90) & 14.36 (3.90) & 14.36 (3.90) & 14.36 (3.90) & 14.36 (3.90)\\
 20  &  & 14.20 (4.38) & 14.05 (4.85) & 13.91 (5.33) & 13.79 (5.81) \\
 10  & & 13.89 (5.40) & 13.53 (6.92) & 13.25 (8.46) & 13.00 (9.99) \\
0  & & 13.20 (8.70) & 12.56 (13.59) & 12.12 (18.51) & 11.77 (23.45)\\
-$\infty$ & & 0 &0 & 0& 0\\
\hline
\end{tabular}
\label{tab_16bits}
\end{table*}
Following the steps in Sec. \ref{sec:entquan} and bounding $\Delta$, we finally arrive at the generalization of Eq.\eqref{eq:wcHminOp},
\begin{equation}
H_{\rm min}(M_{\rm dis}|E)= -\log_2 \max (  c_1, c_2),
\label{eq:wcHminDC}
\end{equation}
Here $c_1 =
\frac{1}{2}\left[\erf\left(\frac{e_{\rm max}+\Delta_{\rm max}-R+3 \delta/2}{\sqrt{2}}\right)+1\right]$ and $c_2 = \erf\left(\frac{\delta}{2 \sqrt{2}}\right)$. The results are tabulated in Tables \ref{tab_8bits} and \ref{tab_16bits}.
\section{Binning of electronic noise - from eavesdropper's perspective}
\label{app:bin}
From Eq.~\eqref{eq:PHybrid}, the discretized electronic noise distribution on the eavesdropper's ADC with dynamical range $R_e$ and digitization $n_e$ is given by
\begin{align}
P_{E_{\rm dis}}(e_j)= \begin{cases}  \int_{-\infty}^{-R_e+\delta_e/2} p_{E}(e) {\rm d}e, &j=j_{\rm min}, \\
\int_{e_j-\delta_e/2}^{e_j+\delta_e/2}p_{E}(e) {\rm d}e ,&  j_{\rm min}< j< j_{\rm max},\\
\int_{R_e-3 \delta_e/2}^{\infty} p_{E}(e) {\rm d}e, &j=j_{\rm max},
\label{eq:PHybridE}
\end{cases}
\end{align} 
where $\delta_e=R_e/2^{n_e-1}$ is the corresponding bin width. In order to achieve the lower bound of the average conditional min-entropy described in Eq.~\eqref{eq:aveHmin}, we imagine that the eavesdropper possesses a device with infinite dynamical ADC range and digitization bits, i.e.\ $R_e \rightarrow \infty$ and $n_e\rightarrow \infty$. As $R_e \rightarrow \infty$, the first and last cases in Eq.~\eqref{eq:PHybridE} can be discarded, and we are left with 
\begin{equation}
P_{E_{\rm dis}}(e_j)= \int_{e_j-\delta_e/2}^{e_j+\delta_e/2}p_{E}(e) {\rm d}e.
\label{eq:PHybridER}
\end{equation}
To evaluate the expression for the discretized conditional probability distribution, we make use of the mean value theorem stated  below:

\textbf{Theorem 1} \textit{Mean value theorem: For any continuous function $f(x)$  on an interval $[a,b]$, there exists some $\bar{x} \in [a,b]$ such that,}
\begin{equation}
\int^b_af(x) {\rm d}x=(b-a)f(\bar{x})
\end{equation}
By invoking Theorem 1, there exists $\bar{e}_j \in [e_j-\delta_e/2,e_j+\delta_e/2]$ such that Eq.~\eqref{eq:PHybridER} can be written as
\begin{equation}
P_{E_{\rm dis}}(e_j)= p_E(\bar{e}_j)\delta_e.
\end{equation}
Substituting this back to Eq.~\eqref{eq:pguess}, we end up with
\begin{equation}
\begin{aligned}
&P_\mathrm{guess}(M_{\rm dis}|E_{\rm dis})\\
&=\left[ \sum_{e_j \in E_{\rm dis}} p_E(\bar{e}_j)\delta_e\max_{m_i \in M_{\rm dis}} P_{M_{\rm dis}|E_{\rm dis}}(m_i|e_j)\right].
\end{aligned}
\label{eq:pguessMV}
\end{equation}
Assuming an infinite binning $\delta_e \rightarrow 0$, the sum becomes an integral,
\begin{equation}
\begin{aligned}
&P_\mathrm{guess}(M_{\rm dis}|E)\\
&=\lim_{\delta_e\rightarrow 0} P_\mathrm{guess}(M_{\rm dis}|E_{\rm dis})\\
&=\left[ \int^\infty_{-\infty} p_E(e)\max_{m_i \in M_{\rm dis}} P_{M_{\rm dis}|E}(m_i|e){\rm d}e\right].
\end{aligned}
\label{eq:pguessMVNA}
\end{equation}
Together with Eq.~\eqref{eq:PcondHybridProb}, we finally arrive at
\begin{equation}
\begin{aligned}
&P_\mathrm{guess}(M_{\rm dis}|E)\\
&=\left[ \int^\infty_{-\infty} p_E(e)\max_{m_i \in M_{\rm dis}} P_{M_{\rm dis}|E}(m_i|e)de \right]\\
&=  \frac{1}{2}\Bigg(\int^{e_1}_{-\infty} P_e(e)\left[1-\erf\left(\frac{e+R-\delta/2}{\sqrt{2}}\right)\right] {\rm d} e\\
&+\left[\erf\left(\frac{e_2}{\sqrt{2} \sigma_E}\right)-\erf\left(\frac{e_1}{\sqrt{2} \sigma_E}\right)\right]\erf\left(\frac{\delta}{2 \sqrt{2}}\right)\\
&+\int^\infty_{e_2} P_e(e)\left[\erf\left(\frac{e-R+3 \delta/2}{\sqrt{2}}\right)+1\right]{\rm d}e \Bigg), 
\end{aligned}
\label{eq:pguessMVNX}
\end{equation}
where $e_1$ and $e_2$ are chosen to satisfy the maximization upon $M_{\rm dis}$ for a given $R$. The optimal $R$ is then determined numerically. This result can be easily generalized to take into account a dc offset with the steps described in Appendix \ref{app:dc}, giving
\begin{equation}
\begin{aligned}
&P_\mathrm{guess}(M_{\rm dis}|E)\\
&=  \frac{1}{2}\Bigg(\int^{e_1}_{-\infty} p_E(e-\Delta)\left[1-\erf\left(\frac{e+\Delta+R-\delta/2}{\sqrt{2}}\right)\right] {\rm d} e\\
&+\left[\erf\left(\frac{e_2-\Delta}{\sqrt{2} \sigma_E}\right)-\erf\left(\frac{e_1-\Delta}{\sqrt{2} \sigma_E}\right)\right]\erf\left(\frac{\delta}{2 \sqrt{2}}\right)\\
&+\int^\infty_{e_2} p_E(e-\Delta)\left[\erf\left(\frac{e+\Delta-R+3 \delta/2}{\sqrt{2}}\right)+1\right]{\rm d}e \Bigg), 
\end{aligned}
\label{eq:pguessMVNXD}
\end{equation}
\begin{figure*}[t]
\includegraphics[width=16cm]{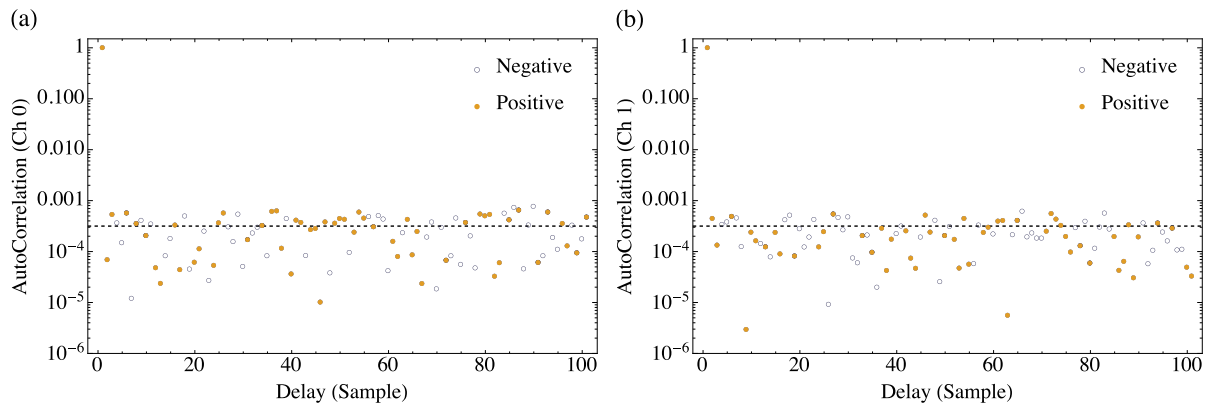}
\caption{Autocorrelation plots of raw samples for (a) channel 0 and (b)
  channel 1 evaluated from a typical record of $10^7$ consecutive samples. \jy{Each sample is 12 bits, where the four most significant bits are discarded from 16-bit raw data.} The low values of autocorrelation between the samples
  are consistent with our raw data being close to independent and identically
  distributed random variables. \SA{Dashed lines show the theoretical
  standard deviation of truly random $10^7$ points}.}
\label{auto}
\end{figure*}
\section{\SA{Upper bound on $H_\text{min}$ for limited laser power}}
\label{app:hmin_limit}
For a finite coherent state $\ket{\alpha}$, the maximum value of
$H_{\rm min}(M_{\rm dis}|E)$ is bounded by the number of photons
available in $\ket{\alpha}$. This limit is attained when the ADC
discretization is fine enough such that events between $n$ and $n+1$
photons at the homodyne output can be distinguished (regardless of the
amount of classical noise). The probability density function
$p_{M|E}(m|e=0)$ is then a probability mass function having support
$(n_1-n_2)\delta_0$ where $n_1$ and $n_2$ are non-negative integers with
a Poissonian distribution with mean
$\abs{\alpha}^2/2$. The normalization constant
$\delta_0=1/\abs{\alpha}$ sets the variance to 1. For large
$\abs{\alpha}$, the distribution $p_{M|E}(m|e=0)$ tends to a
discretised Gaussian distribution with zero mean and unit variance,
\begin{equation}
P_{M_\text{dis}|E}(m|e=0)=
\frac{\delta_0}{\sqrt{2\pi}}\exp\left(-m^2\right)\;,
\end{equation}
for $m\in\left\{0,\pm \delta_0, \pm 2\delta_0, \ldots \right\}$. This
function has a maximum value of $\delta_0/\sqrt{2\pi}$ at
$m=0$.

For an ADC discretization with bin size $\delta$ less
than $\delta_0$ and with range large enough such that the probabilities
of the two end bins given $e$,
$P_{M_{\text{dis}|E}}(m_\text{min}|e)$, and 
 $P_{M_{\text{dis}|E}}(m_\text{max}|e)$ are less than
 $\delta_0/\sqrt{2\pi}$, the most likely
bin given $e$ will have a probability of
$\delta_0/\sqrt{2\pi}$. The min-entropy of this distribution
is then
\begin{align}
&H_{\rm min}(M_{\rm dis}|e) \nonumber\\
&=-\log_2\left[ \max_{m \in M_{\rm dis}} P_{M_{\rm
      dis}|E}(m|e)\right]\nonumber\\
&=-\log_2\left(\frac{\delta_0}{\sqrt{2\pi}} \right)\nonumber\\
&=-\log_2\left(\frac{1}{\sqrt{2\pi}\abs{\alpha}} \right)\;.
\end{align}
Averaging over $e$, this gives the bound to the average conditional
entropy as $\bar{H}_{\rm min}(M_{\rm dis}|E) \leq
-\log_2\left(1/\sqrt{2\pi}\abs{\alpha} \right)$.
\section{Notes on the Leftover Hash Lemma}
\label{sec:lhl}
From an information-theoretic standpoint, the most prominent advantage of universal hashing functions described in Sec.~\ref{sec:RE} is  the randomness of the output guaranteed unconditionally by the leftover hash lemma (LHL). More specifically, LHL states that for any $\varepsilon>0$, if the output of an universal hashing function has length
\begin{equation}
l\leq t -2\log_2(1/\varepsilon), 
\label{eq:lhl}
\end{equation}
where $t$ denotes the (conditional) min-entropy, then the output will be $\varepsilon$-statistically close to a perfectly uniform distribution \cite{Tomamichel2011a}. Moreover, a universal hashing function constructs a strong extractor, where the output string is also independent of the seed of the function \cite{Shaltiel,Ma2013}.

On the other hand, for a strong cryptographic extractor, the output is $\varepsilon'$-computationally indistinguishable from the uniform distribution (see Refs.~\cite{Yvonne2009,Hugo2010} for formal definitions). It is shown in Refs.~\cite{Tomamichel2011a,chevassut2005} that LHL can be generalized to take into account almost universal functions (functions statistically $\xi$-close to being universal hashing functions). \jy{This generalized LHL takes the form of $l=\min(t,\log_2{(1/\xi)})-2s$, where $s$ is an integer related to $\varepsilon'$.} Under suitable parameter constraints and operating modes, an $\varepsilon'$-cryptographic extractor can be treated as a $\xi$-almost universal function, and, hence a strong randomness extractor \cite{chevassut2005,Yvonne2009}. Hence for a cryptographic extractor, it is necessary to sacrifice some bits according to the desired security parameter $e$ to ensure the security and uniformity of the output. 

\section{AES hashing and auto-correlation}
\label{sec:ac}
In our QRNG, randomness extraction is performed with an AES \cite{AEShash} cryptographic hashing
algorithm of 128 bits seeded with a 128-bit secret initialization
vector. Four most significant bits of the 16-bit samples are discarded before randomness
extraction to ensure low autocorrelation among consecutive samples
(Fig.~\ref{auto}) before hashing. The resulting
output is concatenated with partial raw data from the previous run,
forming a 128-bit block for cryptographic hashing. Since a complete
cyptoanalysis of the cryptographic hashing is intricate and is out of
the scope of our work, we simply discard half of the output to ensure
uniformity of the generated random sequence  \cite{Barker2012a}. We further strengthen
our security by renewing the seed of our AES extractor with these
discarded bits. The final real-time throughput of our CV QRNG is
$3.55$ Gbits/s.


\begin{thebibliography}{59}%
\makeatletter
\providecommand \@ifxundefined [1]{%
 \@ifx{#1\undefined}
}%
\providecommand \@ifnum [1]{%
 \ifnum #1\expandafter \@firstoftwo
 \else \expandafter \@secondoftwo
 \fi
}%
\providecommand \@ifx [1]{%
 \ifx #1\expandafter \@firstoftwo
 \else \expandafter \@secondoftwo
 \fi
}%
\providecommand \natexlab [1]{#1}%
\providecommand \enquote  [1]{``#1''}%
\providecommand \bibnamefont  [1]{#1}%
\providecommand \bibfnamefont [1]{#1}%
\providecommand \citenamefont [1]{#1}%
\providecommand \href@noop [0]{\@secondoftwo}%
\providecommand \href [0]{\begingroup \@sanitize@url \@href}%
\providecommand \@href[1]{\@@startlink{#1}\@@href}%
\providecommand \@@href[1]{\endgroup#1\@@endlink}%
\providecommand \@sanitize@url [0]{\catcode `\\12\catcode `\$12\catcode
  `\&12\catcode `\#12\catcode `\^12\catcode `\_12\catcode `\%12\relax}%
\providecommand \@@startlink[1]{}%
\providecommand \@@endlink[0]{}%
\providecommand \url  [0]{\begingroup\@sanitize@url \@url }%
\providecommand \@url [1]{\endgroup\@href {#1}{\urlprefix }}%
\providecommand \urlprefix  [0]{URL }%
\providecommand \Eprint [0]{\href }%
\providecommand \doibase [0]{http://dx.doi.org/}%
\providecommand \selectlanguage [0]{\@gobble}%
\providecommand \bibinfo  [0]{\@secondoftwo}%
\providecommand \bibfield  [0]{\@secondoftwo}%
\providecommand \translation [1]{[#1]}%
\providecommand \BibitemOpen [0]{}%
\providecommand \bibitemStop [0]{}%
\providecommand \bibitemNoStop [0]{.\EOS\space}%
\providecommand \EOS [0]{\spacefactor3000\relax}%
\providecommand \BibitemShut  [1]{\csname bibitem#1\endcsname}%
\let\auto@bib@innerbib\@empty
\bibitem [{\citenamefont {Stip{\v{c}}evi{\'c}}(2011)}]{Stipþeviu2011}%
  \BibitemOpen
  \bibfield  {author} {\bibinfo {author} {\bibfnamefont {Mario}\ \bibnamefont
  {Stip{\v{c}}evi{\'c}}},\ }\bibfield  {title} {\enquote {\bibinfo {title}
  {Quantum random number generators and their use in cryptography},}\ }in\
  \href@noop {} {\emph {\bibinfo {booktitle} {Proceedings of 34th International
  Convention MIPRO}}}\ \bibinfo {pages} {1474–-1479} (\bibinfo {year} {2011})\BibitemShut {NoStop}%
\bibitem [{\citenamefont {Xu}\ \emph {et~al.}(2006)\citenamefont {Xu},
  \citenamefont {Wong}, \citenamefont {Horiuchi},\ and\ \citenamefont
  {Abshire}}]{xu2006compact}%
  \BibitemOpen
  \bibfield  {author} {\bibinfo {author} {\bibfnamefont {P}~\bibnamefont {Xu}},
  \bibinfo {author} {\bibfnamefont {YL}~\bibnamefont {Wong}}, \bibinfo {author}
  {\bibfnamefont {TK}~\bibnamefont {Horiuchi}}, \ and\ \bibinfo {author}
  {\bibfnamefont {PA}~\bibnamefont {Abshire}},\ }\bibfield  {title} {\enquote
  {\bibinfo {title} {Compact floating-gate true random number generator},}\
  }\href@noop {} {\bibfield  {journal} {\bibinfo  {journal} {Electronics
  Letters}\ }\textbf {\bibinfo {volume} {42}},\ \bibinfo {pages} {1346--1347}
  (\bibinfo {year} {2006})}\BibitemShut {NoStop}%
\bibitem [{\citenamefont {Sunar}\ \emph {et~al.}(2007)\citenamefont {Sunar},
  \citenamefont {Martin},\ and\ \citenamefont {Stinson}}]{Sunar2007}%
  \BibitemOpen
  \bibfield  {author} {\bibinfo {author} {\bibfnamefont {Berk}\ \bibnamefont
  {Sunar}}, \bibinfo {author} {\bibfnamefont {William~J}\ \bibnamefont
  {Martin}}, \ and\ \bibinfo {author} {\bibfnamefont {Douglas~R}\ \bibnamefont
  {Stinson}},\ }\bibfield  {title} {\enquote {\bibinfo {title} {A provably
  secure true random number generator with built-in tolerance to active
  attacks},}\ }\href@noop {} {\bibfield  {journal} {\bibinfo  {journal}
  {Computers, IEEE Transactions on}\ }\textbf {\bibinfo {volume} {56}},\
  \bibinfo {pages} {109--119} (\bibinfo {year} {2007})}\BibitemShut {NoStop}%
\bibitem [{\citenamefont {Uchida}\ \emph {et~al.}(2008)\citenamefont {Uchida},
  \citenamefont {Amano}, \citenamefont {Inoue}, \citenamefont {Hirano},
  \citenamefont {Naito}, \citenamefont {Someya}, \citenamefont {Oowada},
  \citenamefont {Kurashige}, \citenamefont {Shiki}, \citenamefont {Yoshimori}
  \emph {et~al.}}]{Uchida2008}%
  \BibitemOpen
  \bibfield  {author} {\bibinfo {author} {\bibfnamefont {Atsushi}\ \bibnamefont
  {Uchida}}, \bibinfo {author} {\bibfnamefont {Kazuya}\ \bibnamefont {Amano}},
  \bibinfo {author} {\bibfnamefont {Masaki}\ \bibnamefont {Inoue}}, \bibinfo
  {author} {\bibfnamefont {Kunihito}\ \bibnamefont {Hirano}}, \bibinfo {author}
  {\bibfnamefont {Sunao}\ \bibnamefont {Naito}}, \bibinfo {author}
  {\bibfnamefont {Hiroyuki}\ \bibnamefont {Someya}}, \bibinfo {author}
  {\bibfnamefont {Isao}\ \bibnamefont {Oowada}}, \bibinfo {author}
  {\bibfnamefont {Takayuki}\ \bibnamefont {Kurashige}}, \bibinfo {author}
  {\bibfnamefont {Masaru}\ \bibnamefont {Shiki}}, \bibinfo {author}
  {\bibfnamefont {Shigeru}\ \bibnamefont {Yoshimori}},  \emph {et~al.},\
  }\bibfield  {title} {\enquote {\bibinfo {title} {Fast physical random bit
  generation with chaotic semiconductor lasers},}\ }\href@noop {} {\bibfield
  {journal} {\bibinfo  {journal} {Nature Photonics}\ }\textbf {\bibinfo
  {volume} {2}},\ \bibinfo {pages} {728--732} (\bibinfo {year}
  {2008})}\BibitemShut {NoStop}%
\bibitem [{\citenamefont {Kanter}\ \emph {et~al.}(2010)\citenamefont {Kanter},
  \citenamefont {Aviad}, \citenamefont {Reidler}, \citenamefont {Cohen},\ and\
  \citenamefont {Rosenbluh}}]{Kanter2009}%
  \BibitemOpen
  \bibfield  {author} {\bibinfo {author} {\bibfnamefont {Ido}\ \bibnamefont
  {Kanter}}, \bibinfo {author} {\bibfnamefont {Yaara}\ \bibnamefont {Aviad}},
  \bibinfo {author} {\bibfnamefont {Igor}\ \bibnamefont {Reidler}}, \bibinfo
  {author} {\bibfnamefont {Elad}\ \bibnamefont {Cohen}}, \ and\ \bibinfo
  {author} {\bibfnamefont {Michael}\ \bibnamefont {Rosenbluh}},\ }\bibfield
  {title} {\enquote {\bibinfo {title} {An optical ultrafast random bit
  generator},}\ }\href@noop {} {\bibfield  {journal} {\bibinfo  {journal}
  {Nature Photonics}\ }\textbf {\bibinfo {volume} {4}},\ \bibinfo {pages}
  {58--61} (\bibinfo {year} {2010})}\BibitemShut {NoStop}%
\bibitem [{\citenamefont {Marangon}\ \emph {et~al.}(2014)\citenamefont
  {Marangon}, \citenamefont {Vallone},\ and\ \citenamefont
  {Villoresi}}]{marangon2014random}%
  \BibitemOpen
  \bibfield  {author} {\bibinfo {author} {\bibfnamefont {Davide~G}\
  \bibnamefont {Marangon}}, \bibinfo {author} {\bibfnamefont {Giuseppe}\
  \bibnamefont {Vallone}}, \ and\ \bibinfo {author} {\bibfnamefont {Paolo}\
  \bibnamefont {Villoresi}},\ }\bibfield  {title} {\enquote {\bibinfo {title}
  {Random bits, true and unbiased, from atmospheric turbulence},}\ }\href@noop
  {} {\bibfield  {journal} {\bibinfo  {journal} {Scientific reports}\ }\textbf
  {\bibinfo {volume} {4}} (\bibinfo {year} {2014})}\BibitemShut {NoStop}%
\bibitem [{\citenamefont {Calude}\ \emph {et~al.}(2010)\citenamefont {Calude},
  \citenamefont {Dinneen}, \citenamefont {Dumitrescu},\ and\ \citenamefont
  {Svozil}}]{Calude2010}%
  \BibitemOpen
  \bibfield  {author} {\bibinfo {author} {\bibfnamefont {Cristian~S}\
  \bibnamefont {Calude}}, \bibinfo {author} {\bibfnamefont {Michael~J}\
  \bibnamefont {Dinneen}}, \bibinfo {author} {\bibfnamefont {Monica}\
  \bibnamefont {Dumitrescu}}, \ and\ \bibinfo {author} {\bibfnamefont {Karl}\
  \bibnamefont {Svozil}},\ }\bibfield  {title} {\enquote {\bibinfo {title}
  {Experimental evidence of quantum randomness incomputability},}\ }\href@noop
  {} {\bibfield  {journal} {\bibinfo  {journal} {Physical Review A}\ }\textbf
  {\bibinfo {volume} {82}},\ \bibinfo {pages} {022102} (\bibinfo {year}
  {2010})}\BibitemShut {NoStop}%
\bibitem [{\citenamefont {Svozil}(2009)}]{Svozil2009}%
  \BibitemOpen
  \bibfield  {author} {\bibinfo {author} {\bibfnamefont {Karl}\ \bibnamefont
  {Svozil}},\ }\bibfield  {title} {\enquote {\bibinfo {title} {Three criteria
  for quantum random-number generators based on beam splitters},}\ }\href@noop
  {} {\bibfield  {journal} {\bibinfo  {journal} {Physical Review A}\ }\textbf
  {\bibinfo {volume} {79}},\ \bibinfo {pages} {054306} (\bibinfo {year}
  {2009})}\BibitemShut {NoStop}%
\bibitem [{\citenamefont {Qi}\ \emph {et~al.}(2010)\citenamefont {Qi},
  \citenamefont {Chi}, \citenamefont {Lo},\ and\ \citenamefont
  {Qian}}]{Qi2010}%
  \BibitemOpen
  \bibfield  {author} {\bibinfo {author} {\bibfnamefont {Bing}\ \bibnamefont
  {Qi}}, \bibinfo {author} {\bibfnamefont {Yue-Meng}\ \bibnamefont {Chi}},
  \bibinfo {author} {\bibfnamefont {Hoi-Kwong}\ \bibnamefont {Lo}}, \ and\
  \bibinfo {author} {\bibfnamefont {Li}~\bibnamefont {Qian}},\ }\bibfield
  {title} {\enquote {\bibinfo {title} {High-speed quantum random number
  generation by measuring phase noise of a single-mode laser},}\ }\href@noop {}
  {\bibfield  {journal} {\bibinfo  {journal} {Optics Letters}\ }\textbf
  {\bibinfo {volume} {35}},\ \bibinfo {pages} {312--314} (\bibinfo {year}
  {2010})}\BibitemShut {NoStop}%
\bibitem [{\citenamefont {Guo}\ \emph {et~al.}(2010)\citenamefont {Guo},
  \citenamefont {Tang}, \citenamefont {Liu},\ and\ \citenamefont
  {Wei}}]{Guo2010}%
  \BibitemOpen
  \bibfield  {author} {\bibinfo {author} {\bibfnamefont {Hong}\ \bibnamefont
  {Guo}}, \bibinfo {author} {\bibfnamefont {Wenzhuo}\ \bibnamefont {Tang}},
  \bibinfo {author} {\bibfnamefont {Yu}~\bibnamefont {Liu}}, \ and\ \bibinfo
  {author} {\bibfnamefont {Wei}\ \bibnamefont {Wei}},\ }\bibfield  {title}
  {\enquote {\bibinfo {title} {Truly random number generation based on
  measurement of phase noise of a laser},}\ }\href@noop {} {\bibfield
  {journal} {\bibinfo  {journal} {Physical Review E}\ }\textbf {\bibinfo
  {volume} {81}},\ \bibinfo {pages} {051137} (\bibinfo {year}
  {2010})}\BibitemShut {NoStop}%
\bibitem [{\citenamefont {Jofre}\ \emph {et~al.}(2011)\citenamefont {Jofre},
  \citenamefont {Curty}, \citenamefont {Steinlechner}, \citenamefont {Anzolin},
  \citenamefont {Torres}, \citenamefont {Mitchell},\ and\ \citenamefont
  {Pruneri}}]{Jofre2011}%
  \BibitemOpen
  \bibfield  {author} {\bibinfo {author} {\bibfnamefont {M}~\bibnamefont
  {Jofre}}, \bibinfo {author} {\bibfnamefont {M}~\bibnamefont {Curty}},
  \bibinfo {author} {\bibfnamefont {F}~\bibnamefont {Steinlechner}}, \bibinfo
  {author} {\bibfnamefont {G}~\bibnamefont {Anzolin}}, \bibinfo {author}
  {\bibfnamefont {JP}~\bibnamefont {Torres}}, \bibinfo {author} {\bibfnamefont
  {MW}~\bibnamefont {Mitchell}}, \ and\ \bibinfo {author} {\bibfnamefont
  {V}~\bibnamefont {Pruneri}},\ }\bibfield  {title} {\enquote {\bibinfo {title}
  {True random numbers from amplified quantum vacuum},}\ }\href@noop {}
  {\bibfield  {journal} {\bibinfo  {journal} {Optics Express}\ }\textbf
  {\bibinfo {volume} {19}},\ \bibinfo {pages} {20665--20672} (\bibinfo {year}
  {2011})}\BibitemShut {NoStop}%
\bibitem [{\citenamefont {Xu}\ \emph {et~al.}(2012)\citenamefont {Xu},
  \citenamefont {Qi}, \citenamefont {Ma}, \citenamefont {Xu}, \citenamefont
  {Zheng},\ and\ \citenamefont {Lo}}]{Xu2012a}%
  \BibitemOpen
  \bibfield  {author} {\bibinfo {author} {\bibfnamefont {Feihu}\ \bibnamefont
  {Xu}}, \bibinfo {author} {\bibfnamefont {Bing}\ \bibnamefont {Qi}}, \bibinfo
  {author} {\bibfnamefont {Xiongfeng}\ \bibnamefont {Ma}}, \bibinfo {author}
  {\bibfnamefont {He}~\bibnamefont {Xu}}, \bibinfo {author} {\bibfnamefont
  {Haoxuan}\ \bibnamefont {Zheng}}, \ and\ \bibinfo {author} {\bibfnamefont
  {Hoi-Kwong}\ \bibnamefont {Lo}},\ }\bibfield  {title} {\enquote {\bibinfo
  {title} {Ultrafast quantum random number generation based on quantum phase
  fluctuations},}\ }\href@noop {} {\bibfield  {journal} {\bibinfo  {journal}
  {Optics Express}\ }\textbf {\bibinfo {volume} {20}},\ \bibinfo {pages}
  {12366--12377} (\bibinfo {year} {2012})}\BibitemShut {NoStop}%
\bibitem [{\citenamefont {Abell{\'a}n}\ \emph {et~al.}(2014)\citenamefont
  {Abell{\'a}n}, \citenamefont {Amaya}, \citenamefont {Jofre}, \citenamefont
  {Curty}, \citenamefont {Ac{\'\i}n}, \citenamefont {Capmany}, \citenamefont
  {Pruneri},\ and\ \citenamefont {Mitchell}}]{Abellan2014}%
  \BibitemOpen
  \bibfield  {author} {\bibinfo {author} {\bibfnamefont {C}~\bibnamefont
  {Abell{\'a}n}}, \bibinfo {author} {\bibfnamefont {W}~\bibnamefont {Amaya}},
  \bibinfo {author} {\bibfnamefont {M}~\bibnamefont {Jofre}}, \bibinfo {author}
  {\bibfnamefont {M}~\bibnamefont {Curty}}, \bibinfo {author} {\bibfnamefont
  {A}~\bibnamefont {Ac{\'\i}n}}, \bibinfo {author} {\bibfnamefont
  {J}~\bibnamefont {Capmany}}, \bibinfo {author} {\bibfnamefont
  {V}~\bibnamefont {Pruneri}}, \ and\ \bibinfo {author} {\bibfnamefont
  {MW}~\bibnamefont {Mitchell}},\ }\bibfield  {title} {\enquote {\bibinfo
  {title} {Ultra-fast quantum randomness generation by accelerated phase
  diffusion in a pulsed laser diode},}\ }\href@noop {} {\bibfield  {journal}
  {\bibinfo  {journal} {Optics Express}\ }\textbf {\bibinfo {volume} {22}},\
  \bibinfo {pages} {1645--1654} (\bibinfo {year} {2014})}\BibitemShut {NoStop}%
\bibitem [{\citenamefont {Stip{\v{c}}evi{\'c}}\ and\ \citenamefont
  {Rogina}(2007)}]{Stipcevic2007}%
  \BibitemOpen
  \bibfield  {author} {\bibinfo {author} {\bibfnamefont {Mario}\ \bibnamefont
  {Stip{\v{c}}evi{\'c}}}\ and\ \bibinfo {author} {\bibfnamefont {B~Medved}\
  \bibnamefont {Rogina}},\ }\bibfield  {title} {\enquote {\bibinfo {title}
  {Quantum random number generator based on photonic emission in
  semiconductors},}\ }\href@noop {} {\bibfield  {journal} {\bibinfo  {journal}
  {Review of scientific instruments}\ }\textbf {\bibinfo {volume} {78}},\
  \bibinfo {pages} {045104} (\bibinfo {year} {2007})}\BibitemShut {NoStop}%
\bibitem [{\citenamefont {Williams}\ \emph {et~al.}(2010)\citenamefont
  {Williams}, \citenamefont {Salevan}, \citenamefont {Li}, \citenamefont
  {Roy},\ and\ \citenamefont {Murphy}}]{Williams2010}%
  \BibitemOpen
  \bibfield  {author} {\bibinfo {author} {\bibfnamefont {Caitlin~RS}\
  \bibnamefont {Williams}}, \bibinfo {author} {\bibfnamefont {Julia~C}\
  \bibnamefont {Salevan}}, \bibinfo {author} {\bibfnamefont {Xiaowen}\
  \bibnamefont {Li}}, \bibinfo {author} {\bibfnamefont {Rajarshi}\ \bibnamefont
  {Roy}}, \ and\ \bibinfo {author} {\bibfnamefont {Thomas~E}\ \bibnamefont
  {Murphy}},\ }\bibfield  {title} {\enquote {\bibinfo {title} {Fast physical
  random number generator using amplified spontaneous emission},}\ }\href@noop
  {} {\bibfield  {journal} {\bibinfo  {journal} {Optics Express}\ }\textbf
  {\bibinfo {volume} {18}},\ \bibinfo {pages} {23584--23597} (\bibinfo {year}
  {2010})}\BibitemShut {NoStop}%
\bibitem [{\citenamefont {Liu}\ \emph {et~al.}(2013)\citenamefont {Liu},
  \citenamefont {Zhu}, \citenamefont {Luo}, \citenamefont {Zhang},\ and\
  \citenamefont {Guo}}]{Liu2013}%
  \BibitemOpen
  \bibfield  {author} {\bibinfo {author} {\bibfnamefont {Y}~\bibnamefont
  {Liu}}, \bibinfo {author} {\bibfnamefont {MY}~\bibnamefont {Zhu}}, \bibinfo
  {author} {\bibfnamefont {B}~\bibnamefont {Luo}}, \bibinfo {author}
  {\bibfnamefont {JW}~\bibnamefont {Zhang}}, \ and\ \bibinfo {author}
  {\bibfnamefont {H}~\bibnamefont {Guo}},\ }\bibfield  {title} {\enquote
  {\bibinfo {title} {Implementation of 1.6 tb s- 1 truly random number
  generation based on a super-luminescent emitting diode},}\ }\href@noop {}
  {\bibfield  {journal} {\bibinfo  {journal} {Laser Physics Letters}\ }\textbf
  {\bibinfo {volume} {10}},\ \bibinfo {pages} {045001} (\bibinfo {year}
  {2013})}\BibitemShut {NoStop}%
\bibitem [{\citenamefont {Wahl}\ \emph {et~al.}(2011)\citenamefont {Wahl},
  \citenamefont {Leifgen}, \citenamefont {Berlin}, \citenamefont
  {R{\"o}hlicke}, \citenamefont {Rahn},\ and\ \citenamefont
  {Benson}}]{Wahl2011}%
  \BibitemOpen
  \bibfield  {author} {\bibinfo {author} {\bibfnamefont {Michael}\ \bibnamefont
  {Wahl}}, \bibinfo {author} {\bibfnamefont {Matthias}\ \bibnamefont
  {Leifgen}}, \bibinfo {author} {\bibfnamefont {Michael}\ \bibnamefont
  {Berlin}}, \bibinfo {author} {\bibfnamefont {Tino}\ \bibnamefont
  {R{\"o}hlicke}}, \bibinfo {author} {\bibfnamefont {Hans-J{\"u}rgen}\
  \bibnamefont {Rahn}}, \ and\ \bibinfo {author} {\bibfnamefont {Oliver}\
  \bibnamefont {Benson}},\ }\bibfield  {title} {\enquote {\bibinfo {title} {An
  ultrafast quantum random number generator with provably bounded output bias
  based on photon arrival time measurements},}\ }\href@noop {} {\bibfield
  {journal} {\bibinfo  {journal} {Applied Physics Letters}\ }\textbf {\bibinfo
  {volume} {98}},\ \bibinfo {pages} {171105} (\bibinfo {year}
  {2011})}\BibitemShut {NoStop}%
\bibitem [{\citenamefont {Wayne}\ and\ \citenamefont
  {Kwiat}(2010)}]{Wayne2010}%
  \BibitemOpen
  \bibfield  {author} {\bibinfo {author} {\bibfnamefont {Michael~A}\
  \bibnamefont {Wayne}}\ and\ \bibinfo {author} {\bibfnamefont {Paul~G}\
  \bibnamefont {Kwiat}},\ }\bibfield  {title} {\enquote {\bibinfo {title}
  {Low-bias high-speed quantum random number generator via shaped optical
  pulses},}\ }\href@noop {} {\bibfield  {journal} {\bibinfo  {journal} {Optics
  Express}\ }\textbf {\bibinfo {volume} {18}},\ \bibinfo {pages} {9351--9357}
  (\bibinfo {year} {2010})}\BibitemShut {NoStop}%
\bibitem [{\citenamefont {Ma}\ \emph {et~al.}(2005)\citenamefont {Ma},
  \citenamefont {Xie},\ and\ \citenamefont {Wu}}]{Ma2005}%
  \BibitemOpen
  \bibfield  {author} {\bibinfo {author} {\bibfnamefont {Hai-Qiang}\
  \bibnamefont {Ma}}, \bibinfo {author} {\bibfnamefont {Yuejian}\ \bibnamefont
  {Xie}}, \ and\ \bibinfo {author} {\bibfnamefont {Ling-An}\ \bibnamefont
  {Wu}},\ }\bibfield  {title} {\enquote {\bibinfo {title} {Random number
  generation based on the time of arrival of single photons},}\ }\href
  {\doibase 10.1364/AO.44.007760} {\bibfield  {journal} {\bibinfo  {journal}
  {Appl. Opt.}\ }\textbf {\bibinfo {volume} {44}},\ \bibinfo {pages}
  {7760--7763} (\bibinfo {year} {2005})}\BibitemShut {NoStop}%
\bibitem [{\citenamefont {Bustard}\ \emph {et~al.}(2013)\citenamefont
  {Bustard}, \citenamefont {England}, \citenamefont {Nunn}, \citenamefont
  {Moffatt}, \citenamefont {Spanner}, \citenamefont {Lausten},\ and\
  \citenamefont {Sussman}}]{Bustard2013}%
  \BibitemOpen
  \bibfield  {author} {\bibinfo {author} {\bibfnamefont {Philip~J}\
  \bibnamefont {Bustard}}, \bibinfo {author} {\bibfnamefont {Duncan~G}\
  \bibnamefont {England}}, \bibinfo {author} {\bibfnamefont {Josh}\
  \bibnamefont {Nunn}}, \bibinfo {author} {\bibfnamefont {Doug}\ \bibnamefont
  {Moffatt}}, \bibinfo {author} {\bibfnamefont {Michael}\ \bibnamefont
  {Spanner}}, \bibinfo {author} {\bibfnamefont {Rune}\ \bibnamefont {Lausten}},
  \ and\ \bibinfo {author} {\bibfnamefont {Benjamin~J}\ \bibnamefont
  {Sussman}},\ }\bibfield  {title} {\enquote {\bibinfo {title} {Quantum random
  bit generation using energy fluctuations in stimulated raman scattering},}\
  }\href@noop {} {\bibfield  {journal} {\bibinfo  {journal} {Optics Express}\
  }\textbf {\bibinfo {volume} {21}},\ \bibinfo {pages} {29350--29357} (\bibinfo
  {year} {2013})}\BibitemShut {NoStop}%
\bibitem [{\citenamefont {Fiorentino}\ \emph {et~al.}(2007)\citenamefont
  {Fiorentino}, \citenamefont {Santori}, \citenamefont {Spillane},
  \citenamefont {Beausoleil},\ and\ \citenamefont {Munro}}]{Munro2007}%
  \BibitemOpen
  \bibfield  {author} {\bibinfo {author} {\bibfnamefont {M}~\bibnamefont
  {Fiorentino}}, \bibinfo {author} {\bibfnamefont {C}~\bibnamefont {Santori}},
  \bibinfo {author} {\bibfnamefont {SM}~\bibnamefont {Spillane}}, \bibinfo
  {author} {\bibfnamefont {RG}~\bibnamefont {Beausoleil}}, \ and\ \bibinfo
  {author} {\bibfnamefont {WJ}~\bibnamefont {Munro}},\ }\bibfield  {title}
  {\enquote {\bibinfo {title} {Secure self-calibrating quantum random-bit
  generator},}\ }\href@noop {} {\bibfield  {journal} {\bibinfo  {journal}
  {Physical Review A}\ }\textbf {\bibinfo {volume} {75}},\ \bibinfo {pages}
  {032334} (\bibinfo {year} {2007})}\BibitemShut {NoStop}%
\bibitem [{\citenamefont {Vallone}\ \emph {et~al.}(2014)\citenamefont
  {Vallone}, \citenamefont {Marangon}, \citenamefont {Tomasin},\ and\
  \citenamefont {Villoresi}}]{Vallone}%
  \BibitemOpen
  \bibfield  {author} {\bibinfo {author} {\bibfnamefont {Giuseppe}\
  \bibnamefont {Vallone}}, \bibinfo {author} {\bibfnamefont {Davide~G}\
  \bibnamefont {Marangon}}, \bibinfo {author} {\bibfnamefont {Marco}\
  \bibnamefont {Tomasin}}, \ and\ \bibinfo {author} {\bibfnamefont {Paolo}\
  \bibnamefont {Villoresi}},\ }\bibfield  {title} {\enquote {\bibinfo {title}
  {Quantum randomness certified by the uncertainty principle},}\ }\href@noop {}
  {\bibfield  {journal} {\bibinfo  {journal} {Physical Review A}\ }\textbf
  {\bibinfo {volume} {90}},\ \bibinfo {pages} {052327} (\bibinfo {year}
  {2014})}\BibitemShut {NoStop}%
\bibitem [{\citenamefont {Symul}\ \emph {et~al.}(2011)\citenamefont {Symul},
  \citenamefont {Assad},\ and\ \citenamefont {Lam}}]{Symul2011}%
  \BibitemOpen
  \bibfield  {author} {\bibinfo {author} {\bibfnamefont {Thomas}\ \bibnamefont
  {Symul}}, \bibinfo {author} {\bibfnamefont {SM}~\bibnamefont {Assad}}, \ and\
  \bibinfo {author} {\bibfnamefont {Ping~K}\ \bibnamefont {Lam}},\ }\bibfield
  {title} {\enquote {\bibinfo {title} {Real time demonstration of high bitrate
  quantum random number generation with coherent laser light},}\ }\href@noop {}
  {\bibfield  {journal} {\bibinfo  {journal} {Applied Physics Letters}\
  }\textbf {\bibinfo {volume} {98}},\ \bibinfo {pages} {231103} (\bibinfo
  {year} {2011})}\BibitemShut {NoStop}%
\bibitem [{\citenamefont {Gabriel}\ \emph {et~al.}(2010)\citenamefont
  {Gabriel}, \citenamefont {Wittmann}, \citenamefont {Sych}, \citenamefont
  {Dong}, \citenamefont {Mauerer}, \citenamefont {Andersen}, \citenamefont
  {Marquardt},\ and\ \citenamefont {Leuchs}}]{Gabriel2010}%
  \BibitemOpen
  \bibfield  {author} {\bibinfo {author} {\bibfnamefont {Christian}\
  \bibnamefont {Gabriel}}, \bibinfo {author} {\bibfnamefont {Christoffer}\
  \bibnamefont {Wittmann}}, \bibinfo {author} {\bibfnamefont {Denis}\
  \bibnamefont {Sych}}, \bibinfo {author} {\bibfnamefont {Ruifang}\
  \bibnamefont {Dong}}, \bibinfo {author} {\bibfnamefont {Wolfgang}\
  \bibnamefont {Mauerer}}, \bibinfo {author} {\bibfnamefont {Ulrik~L}\
  \bibnamefont {Andersen}}, \bibinfo {author} {\bibfnamefont {Christoph}\
  \bibnamefont {Marquardt}}, \ and\ \bibinfo {author} {\bibfnamefont {Gerd}\
  \bibnamefont {Leuchs}},\ }\bibfield  {title} {\enquote {\bibinfo {title} {A
  generator for unique quantum random numbers based on vacuum states},}\
  }\href@noop {} {\bibfield  {journal} {\bibinfo  {journal} {Nature Photonics}\
  }\textbf {\bibinfo {volume} {4}},\ \bibinfo {pages} {711--715} (\bibinfo
  {year} {2010})}\BibitemShut {NoStop}%
\bibitem [{\citenamefont {Sanguinetti}\ \emph {et~al.}(2014)\citenamefont
  {Sanguinetti}, \citenamefont {Martin}, \citenamefont {Zbinden},\ and\
  \citenamefont {Gisin}}]{Sanguinetti}%
  \BibitemOpen
  \bibfield  {author} {\bibinfo {author} {\bibfnamefont {Bruno}\ \bibnamefont
  {Sanguinetti}}, \bibinfo {author} {\bibfnamefont {Anthony}\ \bibnamefont
  {Martin}}, \bibinfo {author} {\bibfnamefont {Hugo}\ \bibnamefont {Zbinden}},
  \ and\ \bibinfo {author} {\bibfnamefont {Nicolas}\ \bibnamefont {Gisin}},\
  }\bibfield  {title} {\enquote {\bibinfo {title} {Quantum random number
  generation on a mobile phone},}\ }\href {\doibase 10.1103/PhysRevX.4.031056}
  {\bibfield  {journal} {\bibinfo  {journal} {Phys. Rev. X}\ }\textbf {\bibinfo
  {volume} {4}},\ \bibinfo {pages} {031056} (\bibinfo {year}
  {2014})}\BibitemShut {NoStop}%
\bibitem [{\citenamefont {Ma}\ \emph {et~al.}(2013)\citenamefont {Ma},
  \citenamefont {Xu}, \citenamefont {Xu}, \citenamefont {Tan}, \citenamefont
  {Qi},\ and\ \citenamefont {Lo}}]{Ma2013}%
  \BibitemOpen
  \bibfield  {author} {\bibinfo {author} {\bibfnamefont {Xiongfeng}\
  \bibnamefont {Ma}}, \bibinfo {author} {\bibfnamefont {Feihu}\ \bibnamefont
  {Xu}}, \bibinfo {author} {\bibfnamefont {He}~\bibnamefont {Xu}}, \bibinfo
  {author} {\bibfnamefont {Xiaoqing}\ \bibnamefont {Tan}}, \bibinfo {author}
  {\bibfnamefont {Bing}\ \bibnamefont {Qi}}, \ and\ \bibinfo {author}
  {\bibfnamefont {Hoi-Kwong}\ \bibnamefont {Lo}},\ }\bibfield  {title}
  {\enquote {\bibinfo {title} {{Postprocessing for quantum random-number
  generators: Entropy evaluation and randomness extraction}},}\ }\href
  {\doibase 10.1103/PhysRevA.87.062327} {\bibfield  {journal} {\bibinfo
  {journal} {Physical Review A}\ }\textbf {\bibinfo {volume} {87}},\ \bibinfo
  {pages} {062327} (\bibinfo {year} {2013})}\BibitemShut {NoStop}%
\bibitem [{\citenamefont {Frauchiger}\ \emph {et~al.}(2013)\citenamefont
  {Frauchiger}, \citenamefont {Renner},\ and\ \citenamefont
  {Troyer}}]{frauchiger2013true}%
  \BibitemOpen
  \bibfield  {author} {\bibinfo {author} {\bibfnamefont {Daniela}\ \bibnamefont
  {Frauchiger}}, \bibinfo {author} {\bibfnamefont {Renato}\ \bibnamefont
  {Renner}}, \ and\ \bibinfo {author} {\bibfnamefont {Matthias}\ \bibnamefont
  {Troyer}},\ }\bibfield  {title} {\enquote {\bibinfo {title} {True randomness
  from realistic quantum devices},}\ }\href@noop {} {\bibfield  {journal}
  {\bibinfo  {journal} {arXiv preprint arXiv:1311.4547}\ } (\bibinfo {year}
  {2013})}\BibitemShut {NoStop}%
\bibitem [{\citenamefont {Oliver}\ \emph {et~al.}(2013)\citenamefont {Oliver},
  \citenamefont {Soriano}, \citenamefont {Sukow},\ and\ \citenamefont
  {Fischer}}]{Oliver2013}%
  \BibitemOpen
  \bibfield  {author} {\bibinfo {author} {\bibfnamefont {Neus}\ \bibnamefont
  {Oliver}}, \bibinfo {author} {\bibfnamefont {Miguel~Cornelles}\ \bibnamefont
  {Soriano}}, \bibinfo {author} {\bibfnamefont {David~W.}\ \bibnamefont
  {Sukow}}, \ and\ \bibinfo {author} {\bibfnamefont {Ingo}\ \bibnamefont
  {Fischer}},\ }\bibfield  {title} {\enquote {\bibinfo {title} {{Fast Random
  Bit Generation Using a Chaotic Laser: Approaching the Information Theoretic
  Limit}},}\ }\href {\doibase 10.1109/JQE.2013.2280917} {\bibfield  {journal}
  {\bibinfo  {journal} {IEEE Journal of Quantum Electronics}\ }\textbf
  {\bibinfo {volume} {49}},\ \bibinfo {pages} {910--918} (\bibinfo {year}
  {2013})}\BibitemShut {NoStop}%
\bibitem [{\citenamefont {Yamazaki}\ and\ \citenamefont
  {Uchida}(2013)}]{Yamazaki2013}%
  \BibitemOpen
  \bibfield  {author} {\bibinfo {author} {\bibfnamefont {T.}~\bibnamefont
  {Yamazaki}}\ and\ \bibinfo {author} {\bibfnamefont {A.}~\bibnamefont
  {Uchida}},\ }\bibfield  {title} {\enquote {\bibinfo {title} {{Performance of
  Random Number Generators Using Noise-Based Superluminescent Diode and
  Chaos-Based Semiconductor Lasers}},}\ }\href {\doibase
  10.1109/JSTQE.2013.2246777} {\bibfield  {journal} {\bibinfo  {journal} {IEEE
  Journal of Selected Topics in Quantum Electronics}\ }\textbf {\bibinfo
  {volume} {19}},\ \bibinfo {pages} {0600309--0600309} (\bibinfo {year}
  {2013})}\BibitemShut {NoStop}%
\bibitem [{\citenamefont {Pironio}\ \emph {et~al.}(2010)\citenamefont
  {Pironio}, \citenamefont {Ac\'{\i}n},\ and\ \citenamefont
  {Massar}}]{Massar2010}%
  \BibitemOpen
  \bibfield  {author} {\bibinfo {author} {\bibfnamefont {S}~\bibnamefont
  {Pironio}}, \bibinfo {author} {\bibfnamefont {A}~\bibnamefont {Ac\'{\i}n}}, \
  and\ \bibinfo {author} {\bibfnamefont {S}~\bibnamefont {Massar}},\ }\bibfield
   {title} {\enquote {\bibinfo {title} {{Random numbers certified by Bell's
  theorem}},}\ }\href {\doibase 10.1038/nature09008} {\bibfield  {journal}
  {\bibinfo  {journal} {Nature}\ }\textbf {\bibinfo {volume} {464}},\ \bibinfo
  {pages} {1021--1024} (\bibinfo {year} {2010})}\BibitemShut {NoStop}%
\bibitem [{\citenamefont {Pironio}\ and\ \citenamefont
  {Massar}(2013)}]{Pironio2013}%
  \BibitemOpen
  \bibfield  {author} {\bibinfo {author} {\bibfnamefont {Stefano}\ \bibnamefont
  {Pironio}}\ and\ \bibinfo {author} {\bibfnamefont {Serge}\ \bibnamefont
  {Massar}},\ }\bibfield  {title} {\enquote {\bibinfo {title} {Security of
  practical private randomness generation},}\ }\href@noop {} {\bibfield
  {journal} {\bibinfo  {journal} {Physical Review A}\ }\textbf {\bibinfo
  {volume} {87}},\ \bibinfo {pages} {012336} (\bibinfo {year}
  {2013})}\BibitemShut {NoStop}%
\bibitem [{\citenamefont {Christensen}\ \emph {et~al.}(2013)\citenamefont
  {Christensen}, \citenamefont {McCusker}, \citenamefont {Altepeter},
  \citenamefont {Calkins}, \citenamefont {Gerrits}, \citenamefont {Lita},
  \citenamefont {Miller}, \citenamefont {Shalm}, \citenamefont {Zhang},
  \citenamefont {Nam} \emph {et~al.}}]{christensen2013detection}%
  \BibitemOpen
  \bibfield  {author} {\bibinfo {author} {\bibfnamefont {BG}~\bibnamefont
  {Christensen}}, \bibinfo {author} {\bibfnamefont {KT}~\bibnamefont
  {McCusker}}, \bibinfo {author} {\bibfnamefont {JB}~\bibnamefont {Altepeter}},
  \bibinfo {author} {\bibfnamefont {B}~\bibnamefont {Calkins}}, \bibinfo
  {author} {\bibfnamefont {T}~\bibnamefont {Gerrits}}, \bibinfo {author}
  {\bibfnamefont {AE}~\bibnamefont {Lita}}, \bibinfo {author} {\bibfnamefont
  {A}~\bibnamefont {Miller}}, \bibinfo {author} {\bibfnamefont
  {LK}~\bibnamefont {Shalm}}, \bibinfo {author} {\bibfnamefont {Y}~\bibnamefont
  {Zhang}}, \bibinfo {author} {\bibfnamefont {SW}~\bibnamefont {Nam}},  \emph
  {et~al.},\ }\bibfield  {title} {\enquote {\bibinfo {title}
  {Detection-loophole-free test of quantum nonlocality, and applications},}\
  }\href@noop {} {\bibfield  {journal} {\bibinfo  {journal} {Physical Review
  Letters}\ }\textbf {\bibinfo {volume} {111}},\ \bibinfo {pages} {130406}
  (\bibinfo {year} {2013})}\BibitemShut {NoStop}%
\bibitem [{\citenamefont {Shaltiel}(2002)}]{Shaltiel}%
  \BibitemOpen
  \bibfield  {author} {\bibinfo {author} {\bibfnamefont {Ronen}\ \bibnamefont
  {Shaltiel}},\ }\bibfield  {title} {\enquote {\bibinfo {title} {Recent
  developments in explicit constructions of extractors},}\ }\href@noop {}
  {\bibfield  {journal} {\bibinfo  {journal} {Bulletin of the EATCS}\ }\textbf
  {\bibinfo {volume} {77}},\ \bibinfo {pages} {67--95} (\bibinfo {year}
  {2002})}\BibitemShut {NoStop}%
\bibitem [{\citenamefont {De}\ \emph {et~al.}(2012)\citenamefont {De},
  \citenamefont {Portmann}, \citenamefont {Vidick},\ and\ \citenamefont
  {Renner}}]{De2012}%
  \BibitemOpen
  \bibfield  {author} {\bibinfo {author} {\bibfnamefont {Anindya}\ \bibnamefont
  {De}}, \bibinfo {author} {\bibfnamefont {Christopher}\ \bibnamefont
  {Portmann}}, \bibinfo {author} {\bibfnamefont {Thomas}\ \bibnamefont
  {Vidick}}, \ and\ \bibinfo {author} {\bibfnamefont {Renato}\ \bibnamefont
  {Renner}},\ }\bibfield  {title} {\enquote {\bibinfo {title} {Trevisan's
  extractor in the presence of quantum side information},}\ }\href@noop {}
  {\bibfield  {journal} {\bibinfo  {journal} {SIAM Journal on Computing}\
  }\textbf {\bibinfo {volume} {41}},\ \bibinfo {pages} {915--940} (\bibinfo
  {year} {2012})}\BibitemShut {NoStop}%
\bibitem [{\citenamefont {Konig}\ \emph {et~al.}(2009)\citenamefont {Konig},
  \citenamefont {Renner},\ and\ \citenamefont {Schaffner}}]{Konig2009}%
  \BibitemOpen
  \bibfield  {author} {\bibinfo {author} {\bibfnamefont {R}~\bibnamefont
  {Konig}}, \bibinfo {author} {\bibfnamefont {Renato}\ \bibnamefont {Renner}},
  \ and\ \bibinfo {author} {\bibfnamefont {Christian}\ \bibnamefont
  {Schaffner}},\ }\bibfield  {title} {\enquote {\bibinfo {title} {The
  operational meaning of min-and max-entropy},}\ }\href@noop {} {\bibfield
  {journal} {\bibinfo  {journal} {Information Theory, IEEE Transactions on}\
  }\textbf {\bibinfo {volume} {55}},\ \bibinfo {pages} {4337--4347} (\bibinfo
  {year} {2009})}\BibitemShut {NoStop}%
\bibitem [{\citenamefont {Mauerer}\ \emph {et~al.}(2012)\citenamefont
  {Mauerer}, \citenamefont {Portmann},\ and\ \citenamefont {Scholz}}]{Mauerer}%
  \BibitemOpen
  \bibfield  {author} {\bibinfo {author} {\bibfnamefont {Wolfgang}\
  \bibnamefont {Mauerer}}, \bibinfo {author} {\bibfnamefont {Christopher}\
  \bibnamefont {Portmann}}, \ and\ \bibinfo {author} {\bibfnamefont
  {Volkher~B}\ \bibnamefont {Scholz}},\ }\bibfield  {title} {\enquote {\bibinfo
  {title} {A modular framework for randomness extraction based on trevisan's
  construction},}\ }\href@noop {} {\bibfield  {journal} {\bibinfo  {journal}
  {arXiv preprint arXiv:1212.0520}\ } (\bibinfo {year} {2012})}\BibitemShut
  {NoStop}%
\bibitem [{\citenamefont {Law}\ \emph {et~al.}(2014)\citenamefont {Law},
  \citenamefont {Bancal}, \citenamefont {Scarani} \emph
  {et~al.}}]{law2014quantum}%
  \BibitemOpen
  \bibfield  {author} {\bibinfo {author} {\bibfnamefont {Yun~Zhi}\ \bibnamefont
  {Law}}, \bibinfo {author} {\bibfnamefont {Jean-Daniel}\ \bibnamefont
  {Bancal}}, \bibinfo {author} {\bibfnamefont {Valerio}\ \bibnamefont
  {Scarani}},  \emph {et~al.},\ }\bibfield  {title} {\enquote {\bibinfo {title}
  {Quantum randomness extraction for various levels of characterization of the
  devices},}\ }\href@noop {} {\bibfield  {journal} {\bibinfo  {journal}
  {Journal of Physics A: Mathematical and Theoretical}\ }\textbf {\bibinfo
  {volume} {47}},\ \bibinfo {pages} {424028} (\bibinfo {year}
  {2014})}\BibitemShut {NoStop}%
\bibitem [{\citenamefont {Dodis}\ \emph {et~al.}(2008)\citenamefont {Dodis},
  \citenamefont {Ostrovsky}, \citenamefont {Reyzin},\ and\ \citenamefont
  {Smith}}]{dodis2008fuzzy}%
  \BibitemOpen
  \bibfield  {author} {\bibinfo {author} {\bibfnamefont {Yevgeniy}\
  \bibnamefont {Dodis}}, \bibinfo {author} {\bibfnamefont {Rafail}\
  \bibnamefont {Ostrovsky}}, \bibinfo {author} {\bibfnamefont {Leonid}\
  \bibnamefont {Reyzin}}, \ and\ \bibinfo {author} {\bibfnamefont {Adam}\
  \bibnamefont {Smith}},\ }\bibfield  {title} {\enquote {\bibinfo {title}
  {Fuzzy extractors: How to generate strong keys from biometrics and other
  noisy data},}\ }\href@noop {} {\bibfield  {journal} {\bibinfo  {journal}
  {SIAM Journal on Computing}\ }\textbf {\bibinfo {volume} {38}},\ \bibinfo
  {pages} {97--139} (\bibinfo {year} {2008})}\BibitemShut {NoStop}%
\bibitem [{\citenamefont {Renner}(2008)}]{renner2008security}%
  \BibitemOpen
  \bibfield  {author} {\bibinfo {author} {\bibfnamefont {Renato}\ \bibnamefont
  {Renner}},\ }\bibfield  {title} {\enquote {\bibinfo {title} {Security of
  quantum key distribution},}\ }\href@noop {} {\bibfield  {journal} {\bibinfo
  {journal} {International Journal of Quantum Information}\ }\textbf {\bibinfo
  {volume} {6}},\ \bibinfo {pages} {1--127} (\bibinfo {year}
  {2008})}\BibitemShut {NoStop}%
\bibitem [{\citenamefont {Shen}\ \emph {et~al.}(2010)\citenamefont {Shen},
  \citenamefont {Tian},\ and\ \citenamefont {Zou}}]{shen2010practical}%
  \BibitemOpen
  \bibfield  {author} {\bibinfo {author} {\bibfnamefont {Yong}\ \bibnamefont
  {Shen}}, \bibinfo {author} {\bibfnamefont {Liang}\ \bibnamefont {Tian}}, \
  and\ \bibinfo {author} {\bibfnamefont {Hongxin}\ \bibnamefont {Zou}},\
  }\bibfield  {title} {\enquote {\bibinfo {title} {Practical quantum random
  number generator based on measuring the shot noise of vacuum states},}\
  }\href@noop {} {\bibfield  {journal} {\bibinfo  {journal} {Physical Review
  A}\ }\textbf {\bibinfo {volume} {81}},\ \bibinfo {pages} {063814} (\bibinfo
  {year} {2010})}\BibitemShut {NoStop}%
\bibitem [{\citenamefont {Durt}\ \emph {et~al.}(2013)\citenamefont {Durt},
  \citenamefont {Belmonte}, \citenamefont {Lamoureux}, \citenamefont
  {Panajotov}, \citenamefont {Van~den Berghe},\ and\ \citenamefont
  {Thienpont}}]{Durt2013}%
  \BibitemOpen
  \bibfield  {author} {\bibinfo {author} {\bibfnamefont {Thomas}\ \bibnamefont
  {Durt}}, \bibinfo {author} {\bibfnamefont {Carlos}\ \bibnamefont {Belmonte}},
  \bibinfo {author} {\bibfnamefont {Louis-Philippe}\ \bibnamefont {Lamoureux}},
  \bibinfo {author} {\bibfnamefont {Krassimir}\ \bibnamefont {Panajotov}},
  \bibinfo {author} {\bibfnamefont {Frederik}\ \bibnamefont {Van~den Berghe}},
  \ and\ \bibinfo {author} {\bibfnamefont {Hugo}\ \bibnamefont {Thienpont}},\
  }\bibfield  {title} {\enquote {\bibinfo {title} {Fast quantum-optical
  random-number generators},}\ }\href@noop {} {\bibfield  {journal} {\bibinfo
  {journal} {Physical Review A}\ }\textbf {\bibinfo {volume} {87}},\ \bibinfo
  {pages} {022339} (\bibinfo {year} {2013})}\BibitemShut {NoStop}%
\bibitem [{\citenamefont {Barak}\ \emph {et~al.}(2003)\citenamefont {Barak},
  \citenamefont {Shaltiel},\ and\ \citenamefont {Tromer}}]{Barak}%
  \BibitemOpen
  \bibfield  {author} {\bibinfo {author} {\bibfnamefont {Boaz}\ \bibnamefont
  {Barak}}, \bibinfo {author} {\bibfnamefont {Ronen}\ \bibnamefont {Shaltiel}},
  \ and\ \bibinfo {author} {\bibfnamefont {Eran}\ \bibnamefont {Tromer}},\
  }\bibfield  {title} {\enquote {\bibinfo {title} {True random number
  generators secure in a changing environment},}\ }\bibfield  {booktitle}
  {\emph {\bibinfo {booktitle} {Cryptographic Hardware and Embedded Systems -
  CHES 2003}},\ }\href@noop {} {\ ,\ \bibinfo {pages} {166--180} (\bibinfo
  {year} {2003})}\BibitemShut {NoStop}%
\bibitem [{\citenamefont {Nie}\ \emph {et~al.}(2014)\citenamefont {Nie},
  \citenamefont {Zhang}, \citenamefont {Zhang}, \citenamefont {Wang},
  \citenamefont {Ma}, \citenamefont {Zhang},\ and\ \citenamefont
  {Pan}}]{Nie2014}%
  \BibitemOpen
  \bibfield  {author} {\bibinfo {author} {\bibfnamefont {You-Qi}\ \bibnamefont
  {Nie}}, \bibinfo {author} {\bibfnamefont {Hong-Fei}\ \bibnamefont {Zhang}},
  \bibinfo {author} {\bibfnamefont {Zhen}\ \bibnamefont {Zhang}}, \bibinfo
  {author} {\bibfnamefont {Jian}\ \bibnamefont {Wang}}, \bibinfo {author}
  {\bibfnamefont {Xiongfeng}\ \bibnamefont {Ma}}, \bibinfo {author}
  {\bibfnamefont {Jun}\ \bibnamefont {Zhang}}, \ and\ \bibinfo {author}
  {\bibfnamefont {Jian-Wei}\ \bibnamefont {Pan}},\ }\bibfield  {title}
  {\enquote {\bibinfo {title} {{Practical and fast quantum random number
  generation based on photon arrival time relative to external reference}},}\
  }\href {\doibase 10.1063/1.4863224} {\bibfield  {journal} {\bibinfo
  {journal} {Applied Physics Letters}\ }\textbf {\bibinfo {volume} {104}},\
  \bibinfo {pages} {051110} (\bibinfo {year} {2014})}\BibitemShut {NoStop}%
\bibitem [{\citenamefont {Barak}\ \emph {et~al.}(2011)\citenamefont {Barak},
  \citenamefont {Dodis}, \citenamefont {Krawczyk}, \citenamefont {Pereira},
  \citenamefont {Pietrzak}, \citenamefont {Standaert},\ and\ \citenamefont
  {Yu}}]{barak2011leftover}%
  \BibitemOpen
  \bibfield  {author} {\bibinfo {author} {\bibfnamefont {Boaz}\ \bibnamefont
  {Barak}}, \bibinfo {author} {\bibfnamefont {Yevgeniy}\ \bibnamefont {Dodis}},
  \bibinfo {author} {\bibfnamefont {Hugo}\ \bibnamefont {Krawczyk}}, \bibinfo
  {author} {\bibfnamefont {Olivier}\ \bibnamefont {Pereira}}, \bibinfo {author}
  {\bibfnamefont {Krzysztof}\ \bibnamefont {Pietrzak}}, \bibinfo {author}
  {\bibfnamefont {Fran{\c{c}}ois-Xavier}\ \bibnamefont {Standaert}}, \ and\
  \bibinfo {author} {\bibfnamefont {Yu}~\bibnamefont {Yu}},\ }\bibfield
  {title} {\enquote {\bibinfo {title} {Leftover hash lemma, revisited},}\ }in\
  \href@noop {} {\emph {\bibinfo {booktitle} {Advances in Cryptology--CRYPTO
  2011}}}\ (\bibinfo  {publisher} {Springer},\ \bibinfo {year} {2011})\ pp.\
  \bibinfo {pages} {1--20}\BibitemShut {NoStop}%
\bibitem [{\citenamefont {Dodis}\ \emph {et~al.}(2004)\citenamefont {Dodis},
  \citenamefont {Gennaro}, \citenamefont {H{\aa}stad}, \citenamefont
  {Krawczyk},\ and\ \citenamefont {Rabin}}]{dodis2004randomness}%
  \BibitemOpen
  \bibfield  {author} {\bibinfo {author} {\bibfnamefont {Yevgeniy}\
  \bibnamefont {Dodis}}, \bibinfo {author} {\bibfnamefont {Rosario}\
  \bibnamefont {Gennaro}}, \bibinfo {author} {\bibfnamefont {Johan}\
  \bibnamefont {H{\aa}stad}}, \bibinfo {author} {\bibfnamefont {Hugo}\
  \bibnamefont {Krawczyk}}, \ and\ \bibinfo {author} {\bibfnamefont {Tal}\
  \bibnamefont {Rabin}},\ }\bibfield  {title} {\enquote {\bibinfo {title}
  {Randomness extraction and key derivation using the cbc, cascade and hmac
  modes},}\ }in\ \href@noop {} {\emph {\bibinfo {booktitle} {Advances in
  Cryptology--CRYPTO 2004}}}\ (\bibinfo {organization} {Springer},\ \bibinfo
  {year} {2004})\ pp.\ \bibinfo {pages} {494--510}\BibitemShut {NoStop}%
\bibitem [{\citenamefont {Krawczyk}(2010)}]{Hugo2010}%
  \BibitemOpen
  \bibfield  {author} {\bibinfo {author} {\bibfnamefont {Hugo}\ \bibnamefont
  {Krawczyk}},\ }\bibfield  {title} {\enquote {\bibinfo {title} {Cryptographic
  extraction and key derivation: The hkdf scheme},}\ }in\ \href@noop {} {\emph
  {\bibinfo {booktitle} {Advances in Cryptology--CRYPTO 2010}}}\ (\bibinfo
  {publisher} {Springer},\ \bibinfo {year} {2010})\ pp.\ \bibinfo {pages}
  {631--648}\BibitemShut {NoStop}%
\bibitem [{\citenamefont {Cliff}\ \emph {et~al.}(2009)\citenamefont {Cliff},
  \citenamefont {Boyd},\ and\ \citenamefont {Nieto}}]{Yvonne2009}%
  \BibitemOpen
  \bibfield  {author} {\bibinfo {author} {\bibfnamefont {Yvonne}\ \bibnamefont
  {Cliff}}, \bibinfo {author} {\bibfnamefont {Colin}\ \bibnamefont {Boyd}}, \
  and\ \bibinfo {author} {\bibfnamefont {Juan~Gonzalez}\ \bibnamefont
  {Nieto}},\ }\bibfield  {title} {\enquote {\bibinfo {title} {How to extract
  and expand randomness: A summary and explanation of existing results},}\ }in\
  \href@noop {} {\emph {\bibinfo {booktitle} {Applied Cryptography and Network
  Security}}}\ (\bibinfo {organization} {Springer},\ \bibinfo {year} {2009})\
  pp.\ \bibinfo {pages} {53--70}\BibitemShut {NoStop}%
\bibitem [{\citenamefont {Chevassut}\ \emph {et~al.}(2006)\citenamefont
  {Chevassut}, \citenamefont {Fouque}, \citenamefont {Gaudry},\ and\
  \citenamefont {Pointcheval}}]{chevassut2005}%
  \BibitemOpen
  \bibfield  {author} {\bibinfo {author} {\bibfnamefont {Olivier}\ \bibnamefont
  {Chevassut}}, \bibinfo {author} {\bibfnamefont {Pierre-Alain}\ \bibnamefont
  {Fouque}}, \bibinfo {author} {\bibfnamefont {Pierrick}\ \bibnamefont
  {Gaudry}}, \ and\ \bibinfo {author} {\bibfnamefont {David}\ \bibnamefont
  {Pointcheval}},\ }\bibfield  {title} {\enquote {\bibinfo {title} {The
  twist-augmented technique for key exchange},}\ }in\ \href@noop {} {\emph
  {\bibinfo {booktitle} {Public Key Cryptography-PKC 2006}}}\ (\bibinfo
  {publisher} {Springer},\ \bibinfo {year} {2006})\ pp.\ \bibinfo {pages}
  {410--426}\BibitemShut {NoStop}%
\bibitem [{\citenamefont {FIPS}(2001)}]{AEShash}%
  \BibitemOpen
  \bibfield  {author} {\bibinfo {author} {\bibfnamefont {PUB}\ \bibnamefont
  {FIPS}},\ }\bibfield  {title} {\enquote {\bibinfo {title} {197: Advanced
  encryption standard (aes)},}\ }\href@noop {} {\bibfield  {journal} {\bibinfo
  {journal} {National Institute of Standards and Technology}\ } (\bibinfo
  {year} {2001})}\BibitemShut {NoStop}%
\bibitem [{\citenamefont {Barker}\ and\ \citenamefont
  {Kelsey}(2012)}]{Barker2012a}%
  \BibitemOpen
  \bibfield  {author} {\bibinfo {author} {\bibfnamefont {Elaine}\ \bibnamefont
  {Barker}}\ and\ \bibinfo {author} {\bibfnamefont {John}\ \bibnamefont
  {Kelsey}},\ }\bibfield  {title} {\enquote {\bibinfo {title} {Recommendation
  for the entropy sources used for random bit generation},}\ }\href@noop {}
  {\bibfield  {journal} {\bibinfo  {journal} {NIST DRAFT Special Publication
  800-90B}\ } (\bibinfo {year} {2012})}\BibitemShut {NoStop}%
\bibitem [{\citenamefont {Rukhin}\ \emph {et~al.}(2010)\citenamefont {Rukhin},
  \citenamefont {Soto}, \citenamefont {Nechvatal}, \citenamefont {Barker},
  \citenamefont {Leigh}, \citenamefont {Levenson}, \citenamefont {Banks},
  \citenamefont {Heckert}, \citenamefont {Dray}, \citenamefont {Vo} \emph
  {et~al.}}]{rukhin2010statistical}%
  \BibitemOpen
  \bibfield  {author} {\bibinfo {author} {\bibfnamefont {Andrew}\ \bibnamefont
  {Rukhin}}, \bibinfo {author} {\bibfnamefont {Juan}\ \bibnamefont {Soto}},
  \bibinfo {author} {\bibfnamefont {James}\ \bibnamefont {Nechvatal}}, \bibinfo
  {author} {\bibfnamefont {Elaine}\ \bibnamefont {Barker}}, \bibinfo {author}
  {\bibfnamefont {Stefan}\ \bibnamefont {Leigh}}, \bibinfo {author}
  {\bibfnamefont {Mark}\ \bibnamefont {Levenson}}, \bibinfo {author}
  {\bibfnamefont {David}\ \bibnamefont {Banks}}, \bibinfo {author}
  {\bibfnamefont {Alan}\ \bibnamefont {Heckert}}, \bibinfo {author}
  {\bibfnamefont {James}\ \bibnamefont {Dray}}, \bibinfo {author}
  {\bibfnamefont {San}\ \bibnamefont {Vo}},  \emph {et~al.},\ }\bibfield
  {title} {\enquote {\bibinfo {title} {Statistical test suite for random and
  pseudorandom number generators for cryptographic applications, nist special
  publication},}\ }\href@noop {} {\  (\bibinfo {year} {2010})}\BibitemShut
  {NoStop}%
\bibitem [{\citenamefont {Marsaglia}(1998)}]{marsaglia1998diehard}%
  \BibitemOpen
  \bibfield  {author} {\bibinfo {author} {\bibfnamefont {Georges}\ \bibnamefont
  {Marsaglia}},\ }\bibfield  {title} {\enquote {\bibinfo {title} {Diehard test
  suite},}\ }\href@noop {} {\bibfield  {journal} {\bibinfo  {journal} {Online:
  http://www. stat. fsu. edu/pub/diehard/. Laste visited}\ }\textbf {\bibinfo
  {volume} {8}},\ \bibinfo {pages} {2014} (\bibinfo {year} {1998})}\BibitemShut
  {NoStop}%
\bibitem [{\citenamefont {Tomamichel}\ \emph {et~al.}(2011)\citenamefont
  {Tomamichel}, \citenamefont {Schaffner}, \citenamefont {Smith},\ and\
  \citenamefont {Renner}}]{Tomamichel2011a}%
  \BibitemOpen
  \bibfield  {author} {\bibinfo {author} {\bibfnamefont {Marco}\ \bibnamefont
  {Tomamichel}}, \bibinfo {author} {\bibfnamefont {Christian}\ \bibnamefont
  {Schaffner}}, \bibinfo {author} {\bibfnamefont {Adam}\ \bibnamefont {Smith}},
  \ and\ \bibinfo {author} {\bibfnamefont {Renato}\ \bibnamefont {Renner}},\
  }\bibfield  {title} {\enquote {\bibinfo {title} {Leftover hashing against
  quantum side information},}\ }\href@noop {} {\bibfield  {journal} {\bibinfo
  {journal} {Information Theory, IEEE Transactions on}\ }\textbf {\bibinfo
  {volume} {57}},\ \bibinfo {pages} {5524--5535} (\bibinfo {year}
  {2011})}\BibitemShut {NoStop}%
\bibitem [{\citenamefont {Gisin}\ and\ \citenamefont
  {Thew}(2010)}]{gisin2010quantum}%
  \BibitemOpen
  \bibfield  {author} {\bibinfo {author} {\bibfnamefont {Nicolas}\ \bibnamefont
  {Gisin}}\ and\ \bibinfo {author} {\bibfnamefont {Robert~Thomas}\ \bibnamefont
  {Thew}},\ }\bibfield  {title} {\enquote {\bibinfo {title} {Quantum
  communication technology},}\ }\href@noop {} {\bibfield  {journal} {\bibinfo
  {journal} {Electronics letters}\ }\textbf {\bibinfo {volume} {46}},\ \bibinfo
  {pages} {965--967} (\bibinfo {year} {2010})}\BibitemShut {NoStop}%
\bibitem [{\citenamefont {Pavel~Lougovski}\ and\ \citenamefont
  {Pooser}(2014)}]{Lougovski2014}%
  \BibitemOpen
  \bibfield  {author} {\bibinfo {author} {\bibfnamefont {Pavel}\ \bibnamefont
  {Pavel~Lougovski}}\ and\ \bibinfo {author} {\bibfnamefont {Raphael}\
  \bibnamefont {Pooser}},\ }\bibfield  {title} {\enquote {\bibinfo {title} {An
  observed-data-consistent approach to the assignment of bit values in a
  quantum random number generator},}\ }\href@noop {} {\bibfield  {journal}
  {\bibinfo  {journal} {arXiv preprint arXiv:1404.5977}\ } (\bibinfo {year}
  {2014})}\BibitemShut {NoStop}%
\bibitem [{\citenamefont {Stip{\v{c}}evi{\'c}}\ and\ \citenamefont
  {Bowers}(2014)}]{stipvcevic2014post}%
  \BibitemOpen
  \bibfield  {author} {\bibinfo {author} {\bibfnamefont {Mario}\ \bibnamefont
  {Stip{\v{c}}evi{\'c}}}\ and\ \bibinfo {author} {\bibfnamefont {John}\
  \bibnamefont {Bowers}},\ }\bibfield  {title} {\enquote {\bibinfo {title}
  {Post-processing free spatio-temporal optical random number generator
  resilient to hardware failure and signal injection attacks},}\ }\href@noop {}
  {\bibfield  {journal} {\bibinfo  {journal} {arXiv preprint arXiv:1410.0724}\
  } (\bibinfo {year} {2014})}\BibitemShut {NoStop}%
\bibitem [{\citenamefont {Lunghi}\ \emph {et~al.}(2014)\citenamefont {Lunghi},
  \citenamefont {Brask}, \citenamefont {Lim}, \citenamefont {Lavigne},
  \citenamefont {Bowles}, \citenamefont {Martin}, \citenamefont {Zbinden},\
  and\ \citenamefont {Brunner}}]{lunghi2014self}%
  \BibitemOpen
  \bibfield  {author} {\bibinfo {author} {\bibfnamefont {Tommaso}\ \bibnamefont
  {Lunghi}}, \bibinfo {author} {\bibfnamefont {Jonatan~Bohr}\ \bibnamefont
  {Brask}}, \bibinfo {author} {\bibfnamefont {Charles Ci~Wen}\ \bibnamefont
  {Lim}}, \bibinfo {author} {\bibfnamefont {Quentin}\ \bibnamefont {Lavigne}},
  \bibinfo {author} {\bibfnamefont {Joseph}\ \bibnamefont {Bowles}}, \bibinfo
  {author} {\bibfnamefont {Anthony}\ \bibnamefont {Martin}}, \bibinfo {author}
  {\bibfnamefont {Hugo}\ \bibnamefont {Zbinden}}, \ and\ \bibinfo {author}
  {\bibfnamefont {Nicolas}\ \bibnamefont {Brunner}},\ }\bibfield  {title}
    {\enquote {\bibinfo {title} {Self-Testing Quantum Random Number Generator},}\ }\href@noop {} {\bibfield  {journal} {\bibinfo
  {journal} {Physical Review Letters}\ }\textbf {\bibinfo {volume} {114}},\ \bibinfo
  {pages} {150501} (\bibinfo {year} {2015})}\BibitemShut {NoStop}%
\bibitem [{\citenamefont {Ca{\~n}as}\ \emph {et~al.}(2014)\citenamefont
  {Ca{\~n}as}, \citenamefont {Cari{\~n}e}, \citenamefont {G{\'o}mez},
  \citenamefont {Barra}, \citenamefont {Cabello}, \citenamefont {Xavier},
  \citenamefont {Lima},\ and\ \citenamefont
  {Paw{\l}owski}}]{canas2014experimental}%
  \BibitemOpen
  \bibfield  {author} {\bibinfo {author} {\bibfnamefont {Gustavo}\ \bibnamefont
  {Ca{\~n}as}}, \bibinfo {author} {\bibfnamefont {Jaime}\ \bibnamefont
  {Cari{\~n}e}}, \bibinfo {author} {\bibfnamefont {Esteban~S}\ \bibnamefont
  {G{\'o}mez}}, \bibinfo {author} {\bibfnamefont {Johanna~F}\ \bibnamefont
  {Barra}}, \bibinfo {author} {\bibfnamefont {Ad{\'a}n}\ \bibnamefont
  {Cabello}}, \bibinfo {author} {\bibfnamefont {Guilherme~B}\ \bibnamefont
  {Xavier}}, \bibinfo {author} {\bibfnamefont {Gustavo}\ \bibnamefont {Lima}},
  \ and\ \bibinfo {author} {\bibfnamefont {Marcin}\ \bibnamefont
  {Paw{\l}owski}},\ }\bibfield  {title} {\enquote {\bibinfo {title}
  {Experimental quantum randomness extraction invulnerable to detection
  loophole attacks},}\ }\href@noop {} {\bibfield  {journal} {\bibinfo
  {journal} {arXiv preprint arXiv:1410.3443}\ } (\bibinfo {year}
  {2014})}\BibitemShut {NoStop}%
\bibitem [{\citenamefont {Mitchell}\ \emph {et~al.}(2015)\citenamefont
  {Mitchell}, \citenamefont {Abellan},\ and\ \citenamefont
  {Amaya}}]{Mitchell2015}%
  \BibitemOpen
  \bibfield  {author} {\bibinfo {author} {\bibfnamefont {Morgan~W.}\
  \bibnamefont {Mitchell}}, \bibinfo {author} {\bibfnamefont {Carlos}\
  \bibnamefont {Abellan}}, \ and\ \bibinfo {author} {\bibfnamefont {Waldimar}\
  \bibnamefont {Amaya}},\ }\bibfield  {title} {\enquote {\bibinfo {title}
  {Strong experimental guarantees in ultrafast quantum random number
  generation},}\ }\href {\doibase 10.1103/PhysRevA.91.012314} {\bibfield
  {journal} {\bibinfo  {journal} {Phys. Rev. A}\ }\textbf {\bibinfo {volume}
  {91}},\ \bibinfo {pages} {012314} (\bibinfo {year} {2015})}\BibitemShut
  {NoStop}%
\end{thebibliography}
\end{document}